\documentclass{article}
\usepackage{cite}
\usepackage{amsmath,amssymb,amsfonts}
\usepackage{algorithmic}
\usepackage{graphicx}
\usepackage{textcomp}
\usepackage{verbatim}
\usepackage{algorithmic}
\usepackage{algorithm}
\usepackage{array}
\usepackage[caption=false,font=normalsize,labelfont=sf,textfont=sf]{subfig}
\usepackage{textcomp}
\usepackage{stfloats}
\usepackage{url}
\usepackage{hyperref}
\usepackage{booktabs}
\usepackage{comment}
\usepackage{bbm}
\usepackage{color, colortbl}
\usepackage{todonotes}
\usepackage[switch]{lineno}
\usepackage[normalem]{ulem}
\usepackage[preprint]{corl_2024} 

\newcommand\undermat[2]{%
  \makebox[0pt][l]{$\smash{\underbrace{\phantom{%
    \begin{matrix}#2\end{matrix}}}_{\text{$#1$}}}$}#2}

\usepackage{mdframed}
\usepackage{ntheorem}
\theoremstyle{nonumberplain}
\newmdtheoremenv[%
  linecolor=red,
  linewidth=2pt,
  rightline=false,
  leftline=false]{figrev}{}

\definecolor{Gray}{gray}{0.9}

\title{Fine-tuning Myoelectric Control through Reinforcement Learning in a Game Environment}

\author{
  Kilian Freitag\\
  Department of Electrical Engineering\\
  Chalmers University of Technology 
  Sweden\\
  \texttt{tamino@chalmers.se} \\
  \And
  Yiannis Karayiannidis\\
  Department of Automatic Control\\
  Lund University 
  Sweden\\
  \texttt{yiannis.karayiannidis@control.lth.se} \\
  \And
  Jan Zbinden\thanks{Shared senior authorship.}\\
  Department of Electrical Engineering\\
  Chalmers University of Technology 
  Sweden\\
  \texttt{zbinden@chalmers.se} \\
  \And
  Rita Laezza\footnotemark[1]\\
  Department of Electrical Engineering\\
  Chalmers University of Technology 
  Sweden\\
  \texttt{laezza@chalmers.se} \\
}

\begin{document}
\maketitle


\begin{abstract}
\textit{Objective:}
Enhancing the reliability of myoelectric controllers that decode motor intent is a pressing challenge in the field of bionic prosthetics. State-of-the-art research has mostly focused on Supervised Learning (SL) techniques to tackle this problem. However, obtaining high-quality labeled data that accurately represents muscle activity during daily usage remains difficult. We investigate the potential of Reinforcement Learning (RL) to further improve the decoding of human motion intent by incorporating usage-based data.
\textit{Methods:} 
The starting point of our method is a SL control policy, pretrained on a static recording of electromyographic (EMG) ground truth data. We then apply RL to fine-tune the pretrained classifier with dynamic EMG data obtained during interaction with a game environment developed for this work. We conducted real-time experiments to evaluate our approach and achieved significant improvements in human-in-the-loop performance.
\textit{Results:}
The method effectively predicts simultaneous finger movements, leading to a two-fold increase in decoding accuracy during gameplay and a 39\% improvement in a separate motion test. 
\textit{Conclusion:} 
By employing RL and incorporating usage-based EMG data during fine-tuning, our method achieves significant improvements in accuracy and robustness.
\textit{Significance:} 
These results showcase the potential of RL for enhancing the reliability of myoelectric controllers, of particular importance for advanced bionic limbs.
See our project page for visual demonstrations: \href{https://sites.google.com/view/bionic-limb-rl}{sites.google.com/view/bionic-limb-rl}.
\end{abstract}



\section{Introduction}
Roughly 58 million people were living with limb amputation worldwide, as of 2017 \cite{mcdonald2021global}. Even though extensive research efforts have gone into developing better prosthetic devices, these are still far from human-level performance. As a consequence, the adoption of prostheses is effortful, and abandonment rates are high, with one of the main factors being a lack of reliable functionality \cite{cordella2016literature, smail2021comfort}. The performance of a prosthesis greatly depends on the accuracy with which the user's motion intent can be decoded to determine the appropriate bionic joint actuation. There are several biological signals that can be used to determine human motion intent, such as electroencephalography (EEG) or electromyography (EMG) measurements. The latter has been shown to be more practical, is more widely adopted \cite{jiang2010myoelectric, li2021gesture}, and is therefore used in this work. 

In myoelectric control, there are two main approaches used to decode motor intent for actuating bionic limbs: (i) direct control and (ii) pattern recognition control. \textbf{Direct control}, involves mapping individual muscle signals — often on a one-to-one basis — to a specific bionic joint. When a mapped muscle contracts and its signal value surpasses a predetermined threshold, the corresponding bionic joint is actuated. This approach is appealing due to its simplicity however, the few signals available from people with amputation are often not easily separable and thus cannot always be mapped in a one-to-one fashion. In practice, direct control frequently involves mapping only two antagonistic muscle groups, with mode switching employed to control more than one Degree of Freedom (DOF), making it impractical to use as the number of DOFs grow. \textbf{Pattern recognition control}, leverages Machine Learning (ML) to train a function approximator which automatically maps biological signals to an intended motion \cite{kuiken2016comparison}. This control method thus bypasses the need to solve the signal separability problem through hand engineering efforts. According to Mereu et al. \cite{mereu2021control}, this approach is preferred over direct control by people with amputation due to its more intuitive usability. 

\begin{figure*}[t]
\begin{center}
\centerline{\includegraphics[width=0.85\linewidth]{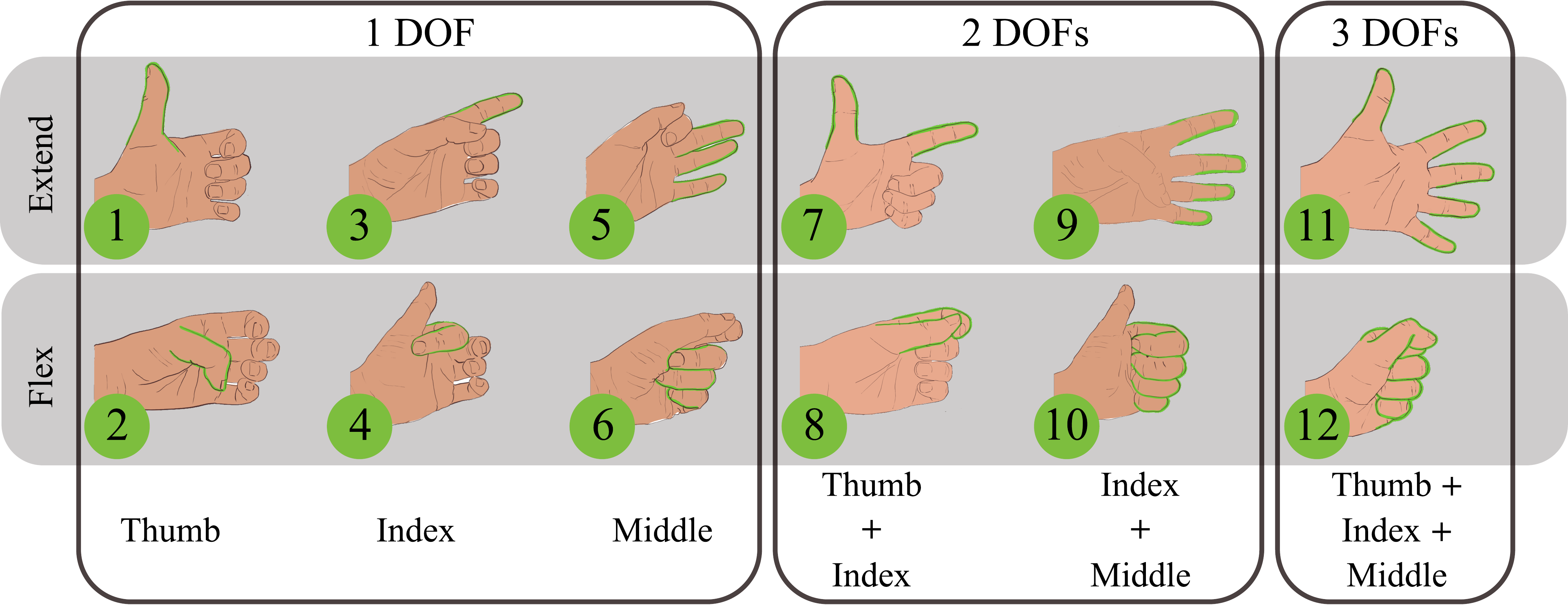}}
\caption{Selected finger movements, grouped by number of simultaneous DOFs. Top row consists of finger extension movements, while bottom row consists of finger flexion movements. Each movement is labeled as $m_i$ with $i=0,\ldots,12$ and with $m_0$ referring to `Rest'.}
\label{fig:movements}
\end{center}
\vspace{-0.4cm}
\end{figure*}

Though pattern recognition control has shown promising results, online usability of ML-based controllers remains limited, especially for hand gestures \cite{li2021gesture}. Many ML approaches have been tested to decode motor intent, from classical algorithms such as support vector machines \cite{oskoei2008support}, to deep learning algorithms using Artificial Neural Networks (ANNs). Most work on pattern recognition control has focused on Supervised Learning (SL) with different ANN architectures, such as feed-forward \cite{hudgins1993new}, recurrent \cite{williams2022recurrent, luu2022artificial}, convolutional \cite{cha2022study} and transformers \cite{godoy2022electromyography}. However, much of this research is confined to offline evaluations, which does not guarantee the same online performance, with a human in the loop \cite{ortiz2015offline}. We, on the other hand, investigate the potential of Reinforcement Learning (RL) methods to increase the online performance of a bionic limb controller, by fine-tuning an initial ANN policy trained via SL.

Other works have attempted to improve ML-based motor intent decoding by combining multiple signals through sensor fusion. Mouchoux et al. \cite{mouchoux2021artificial}, propose integrating inertial units and a camera to obtain context information to improve the performance of a myoelectric controller, in an augmented reality setup.
In contrast, our work aims to maximize the utilization of EMG signals, without additional sensors or hardware, making it a more affordable strategy. Moreover, recent research has explored procedures to improve the EMG signal quality, by implanting electrodes directly in the muscles \cite{vaskov2022surgically} instead of placing them on the surface of the skin, which further encourages pursuing EMG-based strategies. 

A key limitation of ML-based control, is the need to collect labeled EMG data. This typically involves instructing the user to repeatedly contract their muscles to generate activation patterns for each movement that is to be learned during a lengthy recording session. Collecting such ground-truth data becomes increasingly cumbersome as the number of DOFs grows. The problem is exacerbated when simultaneous movements are also considered (e.g. open hand while rotating wrist) since each movement combination needs to be present in the data. Tommasi et al. \cite{tommasi2012improving} proposed using transfer learning, with data from several able-bodied subjects, to reduce the amount of data needed when training a control policy for a new subject. This approach presents challenges when applied to people with amputations due to substantial variations in residual muscle profiles. Furthermore, data from such recording sessions can differ from natural muscle activity during daily usage. This discrepancy between training data and real-life usage can further hinder robust control and thereby impact the functionality of ML-based prostheses. To address these problems, we propose training a control policy within an interactive game setting, which can be more representative of daily-life scenarios. This also helps reduce the length of the initial recording session, since additional data can be collected during gameplay.

In this work, we present an RL-based approach which leverages data collected in an interactive game environment, to close the gap between offline evaluations and online performance. By having users interact with a game through a myoelectric policy, we can get valuable human feedback about its performance. It then becomes possible to iteratively improve the policy towards better online performance directly, rather than only trying to copy recorded behaviors by minimizing the offline error, as in SL approaches. We refer to this iterative improvement of a pretrained policy as fine-tuning, where the purpose is to align the learning objective towards better performance on real usage data, with the human in the loop. While we present a specific case study, our approach can be generalized to any myoelectric control task from which a reward signal can be derived. To validate our methods, we developed a simplified game environment which allows for a quantifiable measure of improvement. The environment is inspired by Guitar Hero which requires precise timing, duration, and motion control, much like everyday movements. The task was developed especially with a simultaneous finger control setting in mind, representative of common movement and grasp patterns encountered in daily life. Our online experimental results demonstrate the efficacy of RL in improving the decoding of motor intent across 15 subjects, with a more than two-fold increase in normalized cumulative reward. The results are further validated through testing on a separate task, also revealing a significant increase~in performance.


\section{Related Work}\label{sec:related_work}
We contextualize our motor intent decoding approach within the broader landscape of advancements in this field, categorizing recent developments into three key areas: a) network architectures, b) data quality and quantity, and c) learning methodologies. We review relevant work in each category, highlighting connections to our proposed method.

\textbf{\textit{a) Neural network architectures:}} There has been a substantial amount of work to investigate the impact of neural network architectures on decoding performance. For instance, Bakircioglu and Özkurt \cite{bakirciouglu2020classification} performed offline tests with a CNN and compared it with a feed-forward fully-connected neural network (FFNN) on 6 different hand movements. Chen et al. \cite{cha2022study} showed that using a Long Short-Term Memory (LSTM) classification head after a Convolutional neural network (CNN) encoder slightly surpassed a fully-connected classification head. Zbinden et al. \cite{zbinden2024deep}, utilizing the same setup as our study, compared fully-connected neural networks with a CNN and Temporal Convolutional Network (TCN) in online experiments. While it is conceivable that more sophisticated network architectures could further enhance motor intent decoding, they also introduce increased complexity. Moreover, we posit that data collection and training methods are at least as crucial as network architecture, yet these areas remain underexplored.

\begin{figure}[t]
\begin{center}
\centerline{\includegraphics[width=.7\columnwidth]{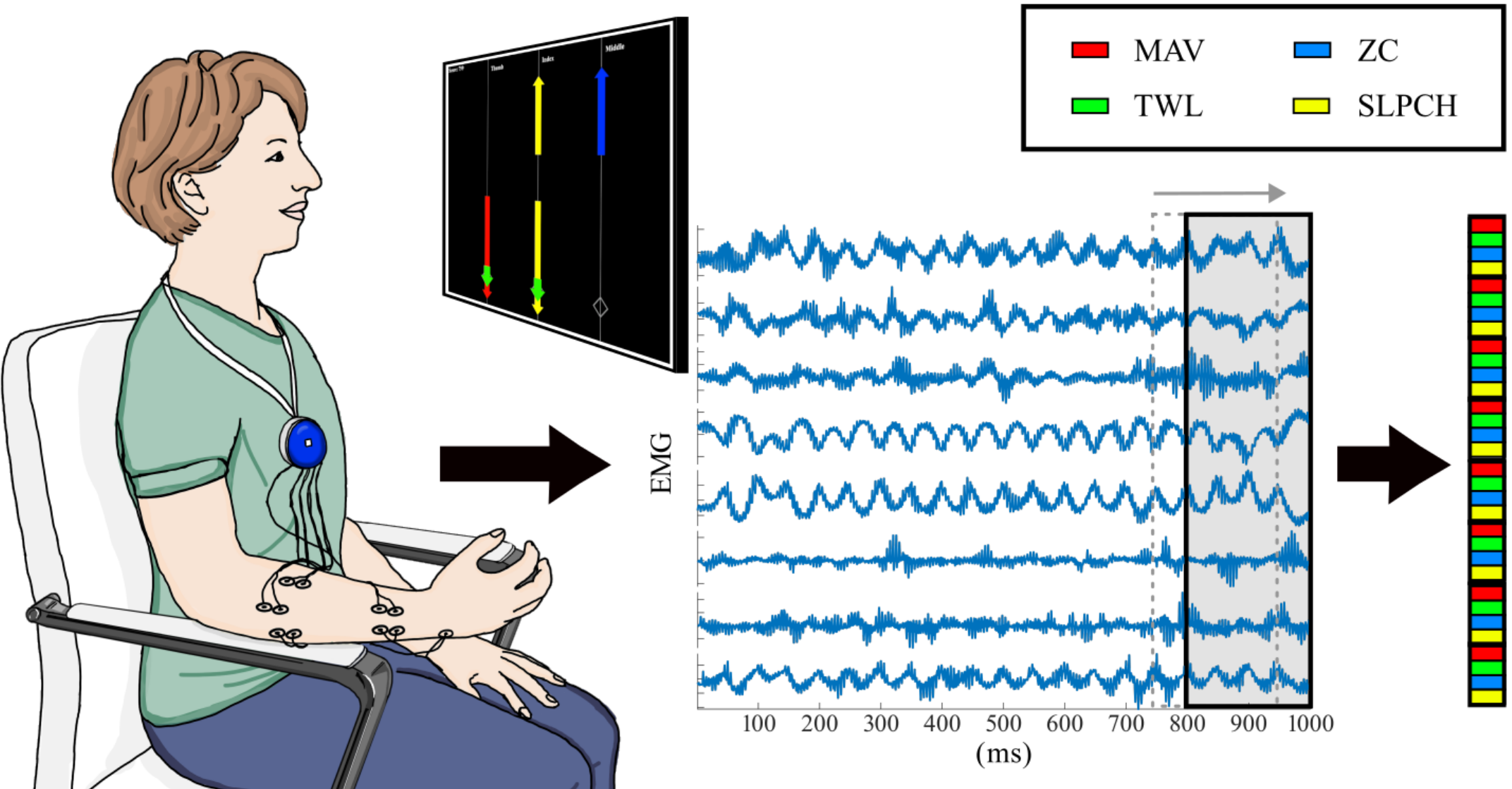}}
\caption{EMG recording setup, with sliding window over 8 input channels from surface electrodes. The Hudgins features \cite{hudgins1993new} — mean absolute value (MAV), waveform length in time-domain (TWL), number of zero crossings (ZC) and slope changes (SLPCH) — are extracted for each recorded channel and stacked in a one-dimensional vector.}
\label{fig:setup}
\end{center}
\end{figure}

\textit{\textbf{b) Data:}} It has been demonstrated in recent studies that increasing the data quantity can lead to generalization \cite{ctrl2024generic}. Nevertheless, work by CTRL-labs at Meta Reality Labs \cite{ctrl2024generic} also showed that additional personalized fine-tuning can improve performance. Moreover, iterative online data collection with fine-tuning has been shown to facilitate learning control tasks in sequence labeling \cite{ross2011reduction} and myoelectric control \cite{gijsberts2014stable}. Building upon the latter findings, we examine how iterative fine-tuning using a gameplay-based approach can improve a policy in an engaging way. In other words, we investigate how data quality impacts learning.

\textit{\textbf{c) Learning methodologies:}} Here, we consider both previous work on transfer learning as it employs a related, yet distinctively different, idea of fine-tuning and, the more closely related work on RL for EMG decoding.

Transfer learning for EMG control has shown promising results in recent work to enable effective usage of additional data. For example, Ketykó et al. \cite{ketyko2019domain} and Tommasi et al. \cite{tommasi2012improving} demonstrated improved generalization across sessions and subjects through fine-tuning on target data. Chen et al. \cite{chen2020hand} demonstrated that transfer learning is also beneficial when the target set contains movements that where not present in the source set. Other work has eliminated the need for labeled source data in domain adaptation \cite{du2017surface, cote2021transferable}, extending the applicability and amount of potential data for the aforementioned methods. We diverge from these approaches by exploring the impact of changing the training environment, and thus the training data distribution, towards tasks that resemble daily usage. Furthermore, we propose using reinforcement learning to facilitate training without requiring labeled data. While we solely investigate the benefit of iterative online fine-tuning with RL, it is possible to integrate transfer learning to leverage usage of additional offline data. We leave this for future work.

Even though EMG classification via SL has been extensively studied, RL methods have received limited exploration. Nevertheless, some promising approaches have emerged in the literature that we want to highlight. Pilarski et al. \cite{pilarski2011online} proposed an RL method that predicts arm movements using EMG and robot state information. The agent learned to match the human arm angle using velocity-based control. They initially guided the learning through a reward based on proximity to the desired angle and, subsequently through sparse human feedback. Similarly, Vasan et al. \cite{vasan2017learning} presented a system that allows for the control of 3 simultaneous DOFs, of a prosthetic limb. This was achieved by employing RL, to train an ANN while recording EMG data of hand movements. Their method was able to predict proportionality for each DOF and was successfully tested on three able-bodied subjects. While these studies demonstrate the feasibility of directly predicting proportional values, our approach focuses on predicting movement intention, which we believe to be more stable and easier to tune for home use. To the best of our knowledge, employing RL for motion intent decoding in this context has not been done before.


\section{Myoelectric Control Scheme}\label{sec:myoelectric}

When designing a myoelectric controller, a key choice is to decide on the specific movements to control. This decision hinges on the participant's amputation level; more proximal limb loss precludes control of movements tied to now-absent muscles. Although our method is broadly applicable, we focus on finger movements corresponding to common grasp patterns, shown in Fig. \ref{fig:movements}. Incongruent articulations (flex + extend combinations) are not selected as there is evidence that natural control of such motions is not simultaneous \cite{rosenbaum2009human}. This selection of finger movements would be feasible for people with trans-radial amputation.
Alternatively, they would be reasonable for patients who have lost the limb more proximally but underwent nerve transfer surgery to create additional myoelectric sites \cite{osborn2021extended, zbinden2023a}.

To acquire EMG signals, eight surface electrode pairs are placed along two rings around the forearm of the participant, with an additional electrode for ground, as illustrated in Fig. \ref{fig:setup}. The electrodes are placed in a bipolar configuration, i.e. for each channel, two adjacent signals are subtracted from one another to reduce noise. The upper ring consists of one electrode pair targeting the \textit{extensor carpi ulnaris} and three additional equally spaced pairs around the proximal side of the forearm. The lower ring includes two electrode pairs targeting the \textit{flexor pollicis longus} and \textit{extensor indicis}, and two equally spaced pairs around the distal side of the forearm. Finally, a single electrode placed on the \textit{ulnar styloid} serves as ground reference. Surface EMG signals are captured at a sampling rate of 1000 Hz and filtered using analog and digital filters. These consist of an analog low-pass filter at 500 Hz, a digital butterworth high-pass filter at 20 Hz, and a second-order notch filter at 50 Hz. The EMG data stream of the eight channels is split with a sliding window approach, into 200 ms windows with 150 ms overlap, equating to an update frequency of 20 Hz.

While there have been works which use raw EMG data to directly train an ANN \cite{zia2018multiday}, such high-dimensional data can make learning more challenging when limited data is available and it increases computational load. A common approach to overcome the problem is to extract a set of four features proposed by Hudgins et al. \cite{hudgins1993new} from the windowed EMG data. The features in question are: mean absolute value (MAV), waveform length in time-domain (TWL), number of zero crossings (ZC) and slope changes (SLPCH). In our myoelectric control scheme, the input to the policy is the stacked vector of these features for each channel, as shown in Fig. \ref{fig:setup}. 

Note how for each DOF (Thumb, Index and Middle, see Fig. \ref{fig:movements}), there are two movements (Flex and Extend). Clinically, sequential myoelectric control is prevalent. This means that for each possible DOF that a prosthesis can actuate, only one can be active at any given time ($m_1$ to $m_6$). In such cases, the ML problem is formulated as a simple classification task. However, since dexterous manipulation requires simultaneous actions, there have been attempts to control multiple DOFs at the same time \cite{jiang2008extracting, zbinden2024deep, ameri2018real}. The most straightforward approach is to treat each movement combination as a new class. Alternatively, we formulate the ML problem as a multi-label classification task. To that end, each movement is encoded into a binary vector $m_i\in \{0, 1\}^{2\cdot{\text{DOF}} \raisebox{0.05ex}{\small+} 1}$, with ${\text{DOF}}=3$. For example, simultaneous thumb and index flexion is encoded as:
\begin{equation}
    m_{8} = \left[
    \begin{array}{cccccccccc}
         \undermat{\text{Thumb}}{0 & 1} & \undermat{\text{Index}}{0 & 1} & \undermat{\text{Middle}}{0 & 0} & \undermat{\text{Rest}}{\! & 0 & \!}
    \end{array} \right] \label{eq:binary_vector}
\end{equation}
\vskip 0.2in

The control policy is chosen to be a relatively small feed-forward, fully-connected (6 hidden layers with ReLU activation, each with 128 neurons) ANN architecture (see \cite{zbinden2024deep} for more details).
Note that our methodology is versatile and can be employed on any network architecture. Considering our primary objective of testing with individuals who have undergone amputation, we strategically selected a network architecture that can be seamlessly integrated into a take-home embedded device, facilitating a smooth transition from research to practical implementation. The output layer has a sigmoid activation function, outputting values between $[0,1]$. At test time, outputs are rounded to be exactly $\{0, 1\}$.

\begin{figure*}[ht]
\begin{center}
\centerline{\includegraphics[width=\linewidth]{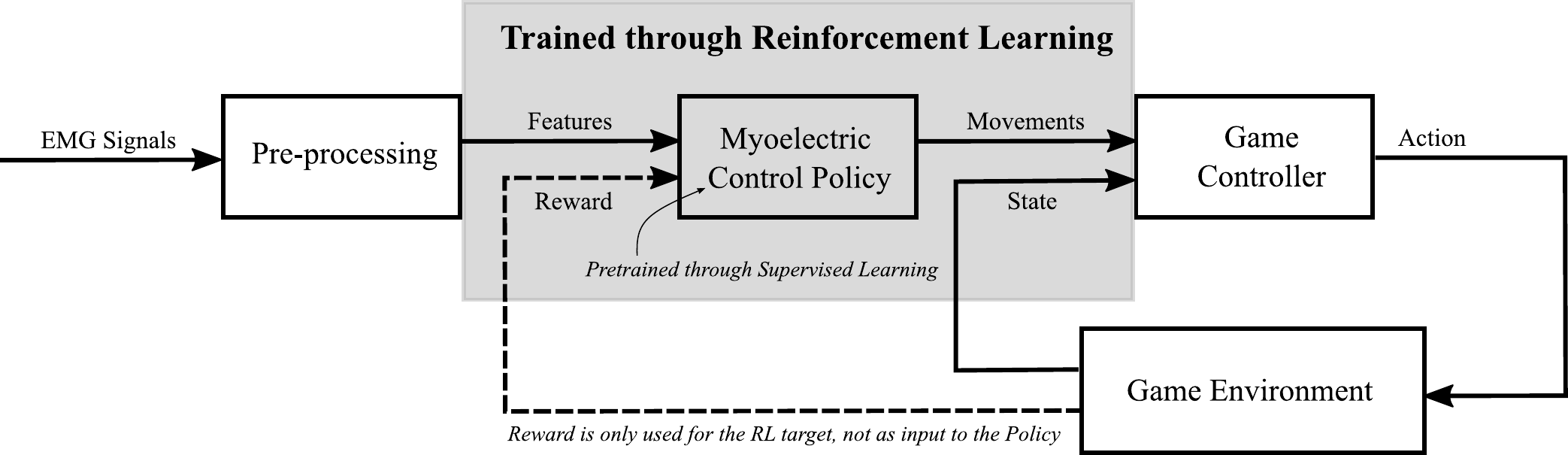}}
\caption{The proposed RL framework consists of obtaining EMG signals from users, that are given to the policy to perform actions in an environment. This environment then gives a reward based on how successful an action was. Note that before interacting with the environment, the policy is pretrained through SL, using data from a recording session. Further note that the reward signal is only used during RL training and not once the policy is deployed.}
\label{fig:rl-framework}
\end{center}
\end{figure*}


\section{Reinforcement Learning}\label{sec:rl}

The idea behind RL is to learn a control policy from trial-and-error while interacting with the environment which provides a reward signal (see Fig. \ref{fig:rl-framework}). More formally, RL problems are formulated as Markov Decision Processes (MDPs). In this work, we consider an episodic MDP setting, defined as a tuple $(\mathcal{S},\mathcal{A},p,r, \gamma)$, where $r: \mathcal{S} \rightarrow \mathbb{R}$ is a reward function and $\gamma\in (0,1]$ is the discount factor. $\mathcal{S}$ and $\mathcal{A}$ are the state and action spaces, respectively. The probability density function $p(s_{t+1} | s_t, a_t)$ represents the probability of transitioning to state $s_{t+1}$, given the current state $s_t$ and action $a_t$, with $s_t, s_{t+1} \in \mathcal{S}$ and $a_t \in \mathcal{A}$. 

We aim to learn an ANN policy $\pi_\theta(s_t)=a_t$, i.e. the actor, parameterized by $\theta$. The long-term objective function is defined by the return, which is the sum of discounted future rewards: $G_t = \sum_{k=t}^{T} \gamma^{k-t} r (s_k, a_k)$. RL algorithms aim to maximize the expected return conditioned on states, i.e. the state-value $V(s),$ or state-action pairs, i.e. the action-value $Q(s,a)$. In deep RL $Q_\phi(s,a)$, also referred to as the critic, is modelled by an ANN parameterized by $\phi$. Typically in actor-critic methods, the actor is updated such that $Q(s,a)$ is maximized:
\begin{equation}
    \theta_{k+1} = \text{arg}\max_{\theta} \mathbb{E} \left[ Q_\phi (s, \pi_{\theta_k} (s) ) \right]
\end{equation}

\subsection{RL with Human in the Loop} \label{subsec:offlinerl}

For our application, the RL agent is a combination of the human and the ANN policy. The former contracts their muscles to execute an intended movement, and the latter maps the recorded EMG signals to the correct movement. Collaboration becomes essential as no single entity can solve tasks independently. 

Online RL presents unique challenges when humans are involved. Firstly, participants having to adapt to policy changes can lead to a less satisfactory experience due to inconsistency between policies. Secondly, the interaction between the human and the environment needs to be real-time, since significant lags between the commands of the human and the game would lead to actions $a_t$ in response to earlier states $s_{t-l}$, where $l$ is the lag. Applying online RL could slow down the game interface, creating such lags. 

An alternative that has been gaining interest in the literature is offline RL \cite{levine2020offline}, which aims to learn from a static dataset, $\mathcal{D}$. This solves the aforementioned problems but brings its own challenges. Because pure offline RL assumes no additional online data collection, it usually cannot reach acceptable online results without further fine-tuning. The recent Advantage Weighted Actor-Critic (AWAC) algorithm proposed by Nair et al. \cite{nair2020awac} aims to accelerate such online fine-tuning with offline datasets. AWAC trains an off-policy critic and an actor with an implicit policy constraint. This leads to the modified policy update, described as:
\begin{equation}
     \theta_{k+1} = \text{arg}\max_\theta \mathbb{E} \left[ \log \pi_\theta  \, \text{exp} \left( \frac{1}{\lambda} A \left(s, \pi_{\theta_k}(s)\right)\right) \right]
\end{equation}
where $ A \left(s, a \right)=Q_\phi(s,a)-V(s)$, is the advantage function and $\lambda$ is the Lagrangian multiplier for the constraint.
By implicitly constraining the actor to stay close to the actions observed in the data, this algorithm was shown to be able to both effectively train offline and continue improving with experience, on real-world robotic problems. Given these particular characteristics, we opted to use AWAC to train the myoelectric control policy, but our method can be combined with other offline RL algorithms.

\subsection{MDP Formulation}\label{subsec:rl_formulation}

The state space $\mathcal{S}\in \mathbb{R}^{32}$ is defined as the stacked vector of 4 features for all 8 channels, as illustrated in Fig. \ref{fig:setup}. The action space $\mathcal{A}\in \{0, 1\}^{7}$ is the binary vector described in equation \eqref{eq:binary_vector}. We select the discount factor to be $\gamma = 0.89$, through a hyperparameter search (see Appendix \ref{sec:hyperparameter} for more details). Furthermore, the reward function is defined as:
\begin{equation}
{r(s_t, a_t)} = \begin{cases}
    1,&{\text{if}}\ \pi(s_t)= a_t^* \land a_t^* \neq m_0 \\ 
    {0,}&{\text{if}}\ \pi(s_t) = a_t^*\land  a_t^* = m_0 \\ 
    -1,&{\text{otherwise.}} 
\end{cases}\label{eq:reward}
\end{equation}

The reward function plays a crucial role in the effectiveness of RL for any given task. Our objective is to train a policy that can accurately predict all movements made by the participant. As described by equation \eqref{eq:reward}, we assign a reward of 1 to correct predictions and a reward of $-1$ to incorrect predictions. 
Furthermore, when no movement is desired. i.e. $a_t^* = m_0$, and the `Rest' class is predicted we assign zero reward. We found this important, since the `Rest' class is over-represented in the data and preliminary tests without zero reward resulted in policies which were biased towards not moving. This reward formulation assumes an ideal participant intention, aligning with the indicated movement, $a_t^*$, determined by the song.

\begin{figure}[t]
\vskip 2mm
\begin{center}
\centerline{\includegraphics[width=\columnwidth]{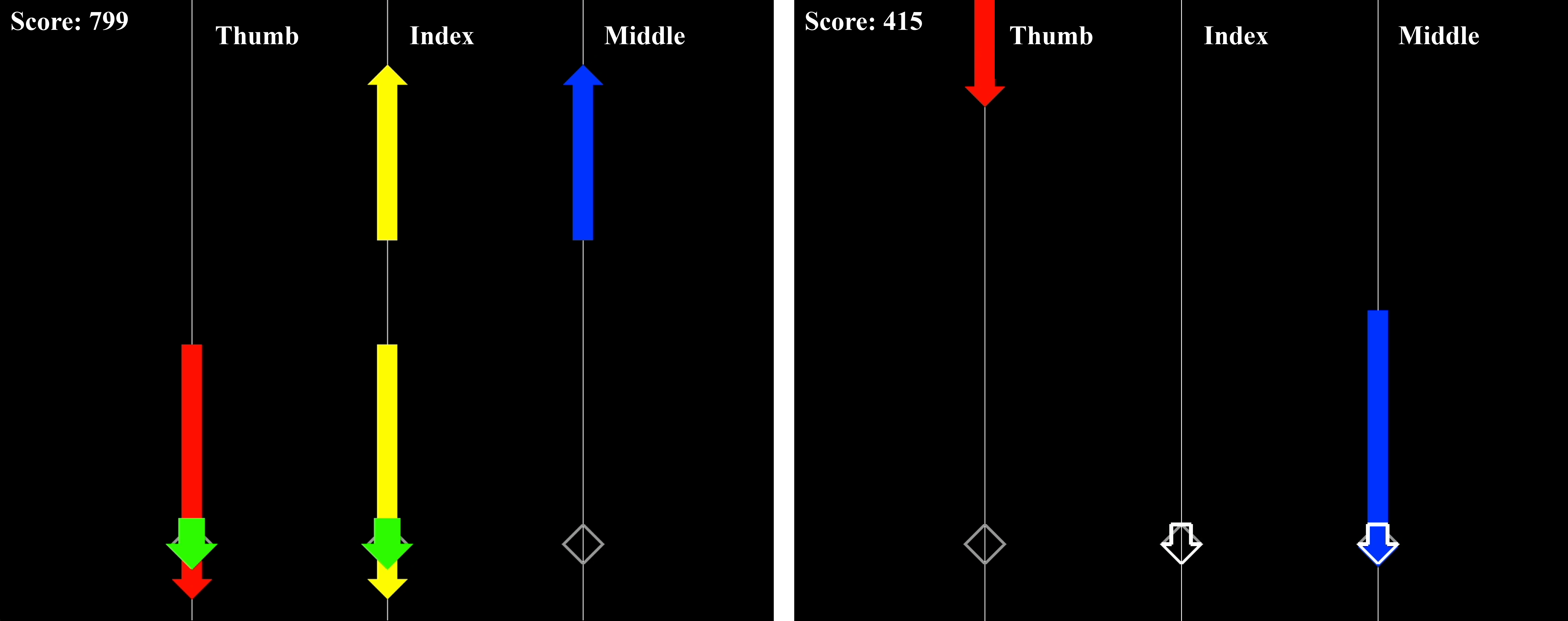}}
\caption{Game interface. Each vertical line refers to a controlled DOF: \textbf{Thumb} (red), \textbf{Index} (yellow) and \textbf{Middle} (blue). The arrows pointing up or down refer to \textbf{extension} and \textbf{flexion}, respectively. Desired movements are shown along the vertical lines, whereas predictions are displayed on the diamonds by short arrows, indicating the direction and DOF that is activated. When the agent executes the desired movement, a green arrow appears over the diamond of the specific DOF (left). Conversely, when the movement is incorrect the arrows are shown in white (right).}
\label{fig:emg-hero}
\end{center}
\end{figure}

The final element of our MDP formulation is the environment itself, whose dynamics are represented by the probability density function $p(s_{t+1} | s_t, a_t)$. To avoid ambiguous results and make it easier to evaluate the performance of the policy-human agent, we designed a task that involves distinct movements at defined times. This simplifies the credit assignment problem (i.e. what action contributes to which reward), when learning the policy. Moreover, in order to make the training process more engaging, we developed a serious game environment similar to Guitar Hero, shown in Fig. \ref{fig:emg-hero}. Note that gamification has been shown to increase participant engagement in previous studies \cite{prahm2018playbionic}. 

The serious game involves synchronizing movements to the beat of a song and displaying them in a way that mimics playing notes. The timing and correctness of the movements can be easily tested in this way. Each song corresponds to one RL episode, with movements lasting for $0.5$, $1.0$, $1.5$, and $2.0$ seconds. These lengths were chosen to align with the time needed for the average prosthetic hand to move, with $2$ seconds being the maximum time elapsed from one extreme to the other. Each movement appears once for each length in one song, thus with 4 repetitions overall. Every repetition uses the same song to allow for a fair comparison. Each episode lasts for $137$ seconds, with $60$ seconds of these filled with notes. For this setting, each episode's undiscounted ($\gamma=1$) return falls within the range $G_0\in[-2740, 1200]$.

However, seeing a negative score could demotivate the participants while interacting with the game. Therefore, we defined a scoring system that is always positive but follows the same trend as the episode's return. The score was displayed on the top left of the user interface to help participants keep track of their performance. In addition to explicit feedback in the form of scores, participants receive implicit feedback through observation of the policy's predictions. This mechanism allows a participant to adapt to policies and to positively reinforce movement patterns that are accurately decoded by a policy, thus enabling more refined muscle activation.


\section{Training Procedure}\label{sec:training}
While Section \ref{sec:rl} presented our RL formulation and interactive game, this section aims to introduce the proposed training procedure designed to experimentally validate our approach. The first step is to execute the traditional procedure for training an SL policy. That is, we perform a recording session where the subject is prompted to repeatedly execute each movement. The resulting raw EMG data is then processed as described in Section \ref{sec:myoelectric} and labeled with the corresponding binary vector $m_i$ for each movement class $i$, to create an initial dataset, $\mathcal{D}_{0} \sim (\mathcal{S},\mathcal{A})$. A policy $\pi_0$ is subsequently trained through SL using $\mathcal{D}_{0}$.

This initial policy is then fine-tuned by playing the serious game described in Section \ref{subsec:rl_formulation}. After the participant finishes the first song, the new EMG data recorded during the game form a dataset $\mathcal{D}_1 \sim (\mathcal{S},\mathcal{A}, \{-1, 0, 1\})$. This new dataset is then used to train a policy $\pi_1$ through offline RL, with policy $\pi_0$ as the starting point. The resulting policy replaces the previous one to play the song again. This process is repeated $n$ times, with each repetition being appended to the dataset i.e. $\mathcal{D}_1 \subset \mathcal{D}_{2} \subset \ldots \subset \mathcal{D}_{n}$. For our experiments, we repeated the procedure $n=8$ times. 

\subsection{SL Pretraining}\label{subsec:sl_pretraining}
To pretrain a policy, each movement (see Fig. \ref{fig:movements}) is recorded 6 times, with a duration of 3 s for each repetition. The first and last 10\% of each recording is discarded to omit transient EMG. Subsequently, the resulting $\mathcal{D}_0$ dataset is used to train the ANN described in Section \ref{sec:myoelectric}. The second and fifth recordings are exclusively reserved as the validation set to determine the best model. Following 500 training epochs, the model with the highest F1 macro score on the validation set is selected as the baseline policy and will be referred to as the pretrained model, i.e. $\pi_0$.

Typically, multi-label classification problems treat each output as the probability of each class being present in the input, by minimizing the sum of binary cross-entropy losses of all classes. Instead, we use a simple RMSE loss, which is more generally used in Behavior Cloning applications \cite{azam2021n} (i.e. SL for control applications).

\subsection{RL Fine-tuning} \label{subsec:rl_fine_tuning}

Based on the proposed RL formulation, we apply the AWAC algorithm introduced in Section \ref{subsec:offlinerl} to train the myoelectric control policy. As one episode provides relatively little data, we include all recorded gameplay data in the replay buffer. In order to enhance the dataset and introduce exploration, we randomize certain actions within dataset $\mathcal{D}_n$. To ensure a smoother playtime experience, the randomization process occurs after each episode, before training rather than in real-time. This randomization happens with a probability $\mathcal{\epsilon} = 0.9$ (see hyperparameter search in Appendix \ref{sec:hyperparameter}) for samples with negative rewards. If triggered, a movement from Fig. \ref{fig:movements} is selected based on a uniform probability distribution. Subsequently, a new reward is calculated for the selected movement.


\section{Experimental Setup} \label{sec:experimental_setup}

This study consists of a preliminary investigation of the potential of our RL-based procedure for improving prosthetic control. Therefore, we first carry out our experiments on able-bodied individuals before moving on to amputees in future work. We recruited 15 participants to perform the experiments, with ages ranging from 21-29. The study protocols were carried out in accordance with the declaration of Helsinki. Signed informed consent was obtained from each participant before conducting the experiments. The study was approved on February 14, 2023, by the Regional Ethical Review Board in Gothenburg (Dnr. 2022-06513-01).

\subsection{Experimental Procedure}

Each experiment was carried out in a single session, lasting between 1-2 hours. The participants were seated comfortably in front of the computer and instructed to rest their arm on the armrest of a chair to avoid any signal disturbances due to motion artifacts or electrode shift. For each participant, we began by pretraining a policy, as described in Section \ref{subsec:sl_pretraining}. During this step, we instructed the user to perform movements at approximately 50-70\% of their maximum voluntary contraction. For all subsequent steps, we did not explicitly specify the contraction level. Notably, data collected from each participant was not used for other participants, as this type of transfer learning is outside the scope of this paper.

Once the SL policy is trained, we let participants play one song without recording data to familiarize themselves with the game and then go through the fine-tuning procedure introduced in Section \ref{sec:training}. In the iterative setting, each repetition has a dual purpose: it first validates the latest policy trained on past data, and then serves as training data itself. Once RL fine-tuning is completed, i.e. after the $n$-th repetition, the participants are asked to play the song one last time using the initial policy, $\pi_0$. This is important to be able to distinguish the impact of RL learning from just human learning. 

Finally, to not only evaluate our approach on gameplay data, which the pretrained policy was not trained on, we also perform a separate online \textit{Motion Test}\cite{kuiken2009targeted}. The participants are prompted to execute each movement for a given number of trials before a timeout, and the classification results are recorded at every time step. Each trial terminates if the correct movement is predicted 40 times (i.e. for 2 seconds) or after a timeout. In this phase, the participants do not get any other instructions than to complete the movements to their best ability. This is performed at the end of each session for both $\pi_0$ and $\pi_8$, where either $\pi_0$ or $\pi_8$ is picked at random, to compare the SL policy with the final RL policy. The participants are not informed of which policy was selected. Although we do not anticipate any confounders, randomly selecting the testing order should mitigate potential effects arising from motion artifacts, electrode shifts, or fatigue. We have the participants repeat each movement 3 times in random order, and set a timeout of 10 seconds. See our project page for visual examples of the experiments: \href{https://sites.google.com/view/bionic-limb-rl}{sites.google.com/view/bionic-limb-rl}.

\subsection{Evaluation Metrics} \label{sec:evaluation}

The most straightforward way to measure gameplay performance differences between policies, is to record the return of each episode/repetition. The undiscounted return (cumulative reward) is normalized based on the range of the game score, presented in Section \ref{subsec:rl_formulation}:
\begin{equation}
    \bar{G}_0 = \frac{\sum_{t=0}^{T} r (s_t, a_t) + 2740}{3940} 
\end{equation}
where $t$ and $T$ are the current and terminal state's indices, respectively. Note that the normalized return is only used during gameplay and not considered during a \textit{Motion Test}.

To further evaluate a policy's capability of decoding motor intent, we also calculate the EMR and the F1 macro scores. In the context of ML classification tasks, the EMR corresponds to the classification accuracy, however, in a multi-label setting, this terminology can be ambiguous since there can be partially correct classifications. For this reason, we refer to the EMR instead, which measures the proportion of correct predictions out of all predictions made by the classifier. When computing the EMR, even partially correct classifications are considered completely incorrect, emphasizing the requirement for precise and accurate predictions. The EMR is defined as:
\begin{equation}
    \text{EMR} = \frac{1}{T} \sum_{t=1}^{T} \mathbbm{1} ( a_t = a_t^*)
\end{equation}
where $\mathbbm{1}$ is the indicator function, $a_t$ the movement predicted and $a_t^*$ the correct movement at time $t$, which is determined by the song being played.

Nevertheless, if some but not all of the target labels are predicted, one could argue that this is more accurate than a case in which none of the target labels are predicted. Therefore we also consider the F1 score for evaluation, which offers a class-wise assessment of performance and is a commonly used indicator for multi-label classification tasks \cite{yang1999evaluation}. The F1 score for each class, $i=1,\ldots,N$, is computed as:
\begin{equation}
    \text{F1}_i= \frac{\text{TP}_i}{\text{TP}_i + 0.5 (\text{FP}_i+\text{FN}_i)}
\end{equation}
where TP denotes True Positives, FP represents False Positives, and FN corresponds to False Negatives. The F1 macro score is obtained by taking the macro average:
\begin{equation}
    \text{F1}_{\text{macro}}= \frac{1}{N} \sum_{i=1}^{N} \text{F1}_i
\end{equation}
Together, these evaluation metrics provide comprehensive insights into the effectiveness of the policy for multi-label classification tasks.

\begin{figure}[t]
\begin{center}
\centerline{\includegraphics[width=.7\columnwidth]{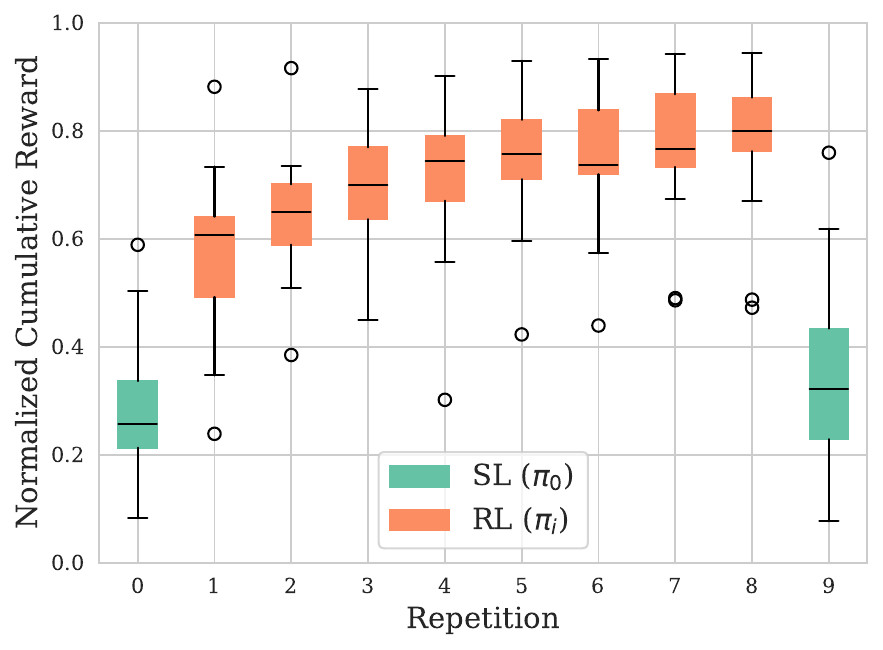}}
\caption{Normalized average cumulative reward over all subjects for RL training repetitions. The first and last repetition is done with the initial pretrained SL policy $\pi_0$, so RL training is only done between repetitions 0 and 8 using the most recent policy $\pi_i$. For one participant our method did not seem to find patterns and thus performed poorly. The lower outliers in most repetitions belong to this participant. The outliers in repetition 7 and 8 belong to another participant who's initial policy was underperforming, but RL did increase performance. Additionally, there are some over-performing outliers. For one participant pretraining worked exceptionally well as seen in repetition 0 and 9. The outliers in repetition 1 and 2 belong to another participant where RL training improved motor decoding faster than usual. The normalized average cumulative reward significantly increases in most repetitions. The differences between repetitions 3-4, 5-6, and 7-8 were not statistically significant. }
\label{fig:rl-cumulative-reward}
\end{center}
\end{figure}

\begin{figure*}[t]
\begin{center}
\centerline{\includegraphics[width=0.95\linewidth]{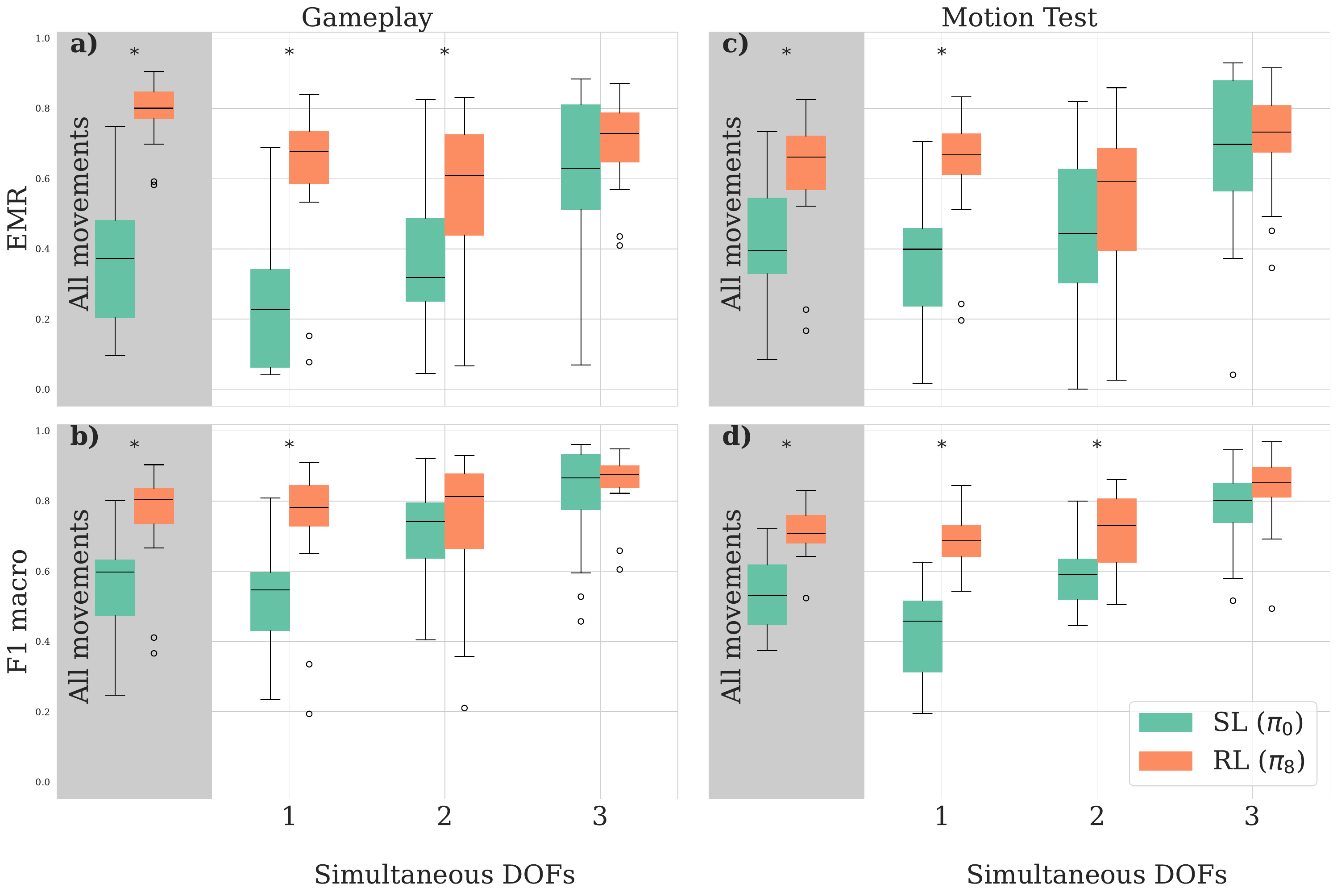}}
\caption{EMR and F1 macro for all movements in Fig. \ref{fig:movements} and per DOF for both gameplay and \textit{Motion Test}. Note that all movements also include $m_0$, 'Rest', which is not included in the box plots per DOF. For gameplay, repetition 8 with policy $\pi_8$ and repetition 9 with $\pi_0$ are compared. Every measure in all scenarios improves with RL, especially for movements that performed poorly before, namely 1 DOF classes. Similar to before, the lower outliers in RL belong to the same individuals for whom RL failed to enhance performance or showed poor performance under $\pi_0$. Columns marked with a star (*) demonstrated a statistically significant improvement. }
\label{fig:results-per-dof}
\end{center} 
\end{figure*}

\subsection{Model Selection}\label{subsubsec:model_selection} 

When performing our experiments there is a need to select the policy to be tested with the participants. This is due to the stochasticity of RL algorithms which leads to non-monotonic improvement of the learned policies. Indeed, determining if under- or over-fitting occurs is still an open problem \cite{kumar2022workflow}. Thus, in each training repetition, the best result may not be the final policy. While the best procedure for model selection is to have the participant test each policy online, we want to minimize the amount of tests one must carry out, to prevent fatigue. Consequently, intermediate policies are tested offline based on all recorded game data so far. During training, the song is simulated every 10 gradient steps, using the recoded data as input for each intermediate model to assess improvements in episodic return. For each repetition, we select the policy with the highest simulated episode return after 2000 gradient steps. This showed superior performance compared to training for a predefined length and selecting the final policy.

\subsection{Statistical Analysis}
After completing the experimental procedure with all participants, we conducted a statistical analysis to evaluate the significance of our findings. Since our results are paired (for each participant), we utilized the Wilcoxon signed-rank test \cite{wilcoxon1992individual}. This test is appropriate for our setting as the metrics should come from the same distribution, though they are not necessarily normally distributed.

The metrics introduced in Section \ref{sec:evaluation} were evaluated for significance in both gameplay and \textit{Motion Test} data. If $\text{p-value} < 0.05$, we considered a change to be significant, which is in accordance with common literature standards.


\section{Results \& Discussion}\label{sec:results}

The average gameplay return for each repetition across all participants from our experiments can be found in Fig. \ref{fig:rl-cumulative-reward}. In addition, Fig. \ref{fig:results-per-dof}, shows the EMR and F1 macro results for both gameplay and \textit{Motion Test} data, separated by number of simultaneous DOFs. For an overall comparison between the pretrained SL policy and the final RL policy, numerical results are summarized in Table \ref{tab:combined_results}.

As can be seen in Fig. \ref{fig:rl-cumulative-reward}, the normalized return increased across nearly all repetitions. We found that it increased by more than 2.5 times (from $0.293\pm 0.136$ to $0.775\pm 0.137$) from the initial SL policy, $\pi_0$, to the last RL repetition, $\pi_8$.  While results appear to plateau, the observed increase between repetitions 7 and 8 implies that additional training could yield further benefits. Moreover, while the normalized average return of the initial SL policy, $\pi_0$, in the final repetition is higher than in repetition 0 ($0.348\pm 0.188$ compared to $0.293\pm 0.136$), it is still less than half of that from the final RL policy, $\pi_8$. This indicates that the RL approach had a significant impact in improving performance ($\text{p-value} = 6.1\times10^{-5}$), despite the fact that participants also learned to play the game better. These results were corroborated by the EMR more than doubling (from $0.36\pm 0.19$ to $0.78\pm 0.09$ with $\text{p-value} = 6.1\times10^{-5}$, see Fig. \ref{fig:results-per-dof}a) and a 40\% increase in F1 scores (from $0.55\pm 0.14$ to $0.75\pm 0.16$ with $\text{p-value} = 6.1\times10^{-5}$, see Fig. \ref{fig:results-per-dof}b). 

In order to understand if the aforementioned improvements were still observed outside the game environment, an additional comparison was performed based on the final \textit{Motion Test} results. Indeed, the mean EMR of the RL policy $\pi_8$ also improved significantly ($\text{p-value} = 1.83\times10^{-4}$) by 39\% (from $0.43\pm 0.18$ to $0.60\pm 0.18$, see Fig. \ref{fig:results-per-dof}c) compared to the SL policy, $\pi_0$. Moreover, significant improvements were observed in the F1 macro score ($\text{p-value} = 6.1\times10^{-5}$), with an increase of 35\% (from $0.53\pm 0.10$ to $0.71\pm 0.07$, see Fig. \ref{fig:results-per-dof}d).

However, the EMR improvements in the \textit{Motion Test} were slightly lower, when compared with gameplay data. 
One possible reason for this discrepancy could be the difference in the participant's focus between the two settings. During gameplay, the primary focus is on the game, requiring correct timing, duration and execution of movements, while in the \textit{Motion Test}, the participant's sole focus is on how to execute one movement at each prompt. This shift in attention may result in altered muscle activation, potentially contributing to the variations in observed metrics. Indeed the procedure of the \textit{Motion Test} is more similar to how the initial recording session is carried out. Nevertheless, it is evident that $\pi_8$ still clearly outperforms $\pi_0$ across all measures.

\begin{table}[h!]
\caption{\textup{Comparison of classification results during gameplay and \textit{Motion Test}, using policy $\pi_8$ (in repetition 8) and the initial policy, $\pi_0$ (in repetition 9). Results are presented as the mean value and standard deviation, across all participants. Results are rounded to two decimal places. }}
\label{tab:combined_results}
\vspace{-0.4cm}
\begin{center}
\begin{small}
\begin{sc}
\begin{tabular}{ccccc}
            \toprule 
           & \multicolumn{2}{c}{Gameplay} & \multicolumn{2}{c}{Motion Test} \\
           \midrule
           & EMR        & $\text{F1}_{\text{\textup{macro}}}$        & EMR          & $\text{F1}_{\text{\textup{macro}}}$          \\
$\pi_0$ & $0.36\pm 0.19$ & $0.55\pm 0.14$ & $0.43\pm 0.18$ & $0.53\pm 0.10$ \\
$\pi_8$ & $\mathbf{0.78}\pm 0.09$ & $\mathbf{0.75}\pm 0.16$  & $\mathbf{0.60}\pm 0.18$ & $\mathbf{0.71}\pm 0.07$ \\
\bottomrule
\end{tabular}
\end{sc}
\end{small}
\end{center}
\vspace{-0.4cm}
\end{table}

\subsection{How do our findings relate to the broader landscape?}
Here, we aim to contextualize our results in relation to recent advances in EMG signal classification.

When focusing on the network architecture, Bakircioglu and Özkurt \cite{bakirciouglu2020classification} outperformed a FFNN with a CNN by $4\%$ for 6 different hand movements in offline tests. Chen et al. \cite{cha2022study} showed that an LSTM classification head after a CNN encoder slightly surpassed a fully connected classification head (97.34 \% vs 93.32 \% accuracy). Similarly, Zbinden et al. \cite{zbinden2024deep} showed that replacing a six-layer FFNN with a CNN (which surpassed a TCN) resulted in an F1 macro score enhancement of 5.7\% in an online \textit{Motion Test}. Our findings, which demonstrate improvements of up to 250\% in gameplay and 35\% of F1 macro score in a \textit{Motion Test} using the same six-layer fully-connected network as in \cite{zbinden2024deep} underscore the significance of data collection and learning method instead of exclusively focusing on the network architecture.

Chen et al. \cite{chen2020hand} demonstrated substantial gains in accuracy using their proposed transfer learning approach. Specifically, they achieved an enhancement of 67.8\% when employing a CNN-only architecture and 28.4\% with a CNN+LSTM, compared to  plain SL training, in offline evaluations for scenarios where the target labels differ from the source labels. Furthermore, Hannius et al. \cite{hannius2024towards} demonstrate a 9 \% improvement in online decoding accuracy by employing Sliced Wasserstein Discrepancy in a limb position domain shift problem. While transfer learning mainly aims to improve generalization by using additional offline data, our results show that the problem of distributional shift between training and deployment can also be tackled during fine-tuning, as evidenced by our competitive improvements. 
It is important to note that our method does not replace transfer learning, instead we show that carefully designed fine-tuning can significantly increase decoding performance as well. Further note that our results are not directly comparable to those by Vasan and Pilarski \cite{vasan2017learning} as in their RL approach they estimate joint velocities, whereas we focus on motion classification.

A limitation of our study is that we are unable to independently assess the distinct contributions of our game environment and RL training procedure. We leave it as an open issue for future research to explore the independent effects of each component.

\begin{figure}[t]
\begin{center}
\centerline{\includegraphics[width=0.7\columnwidth]{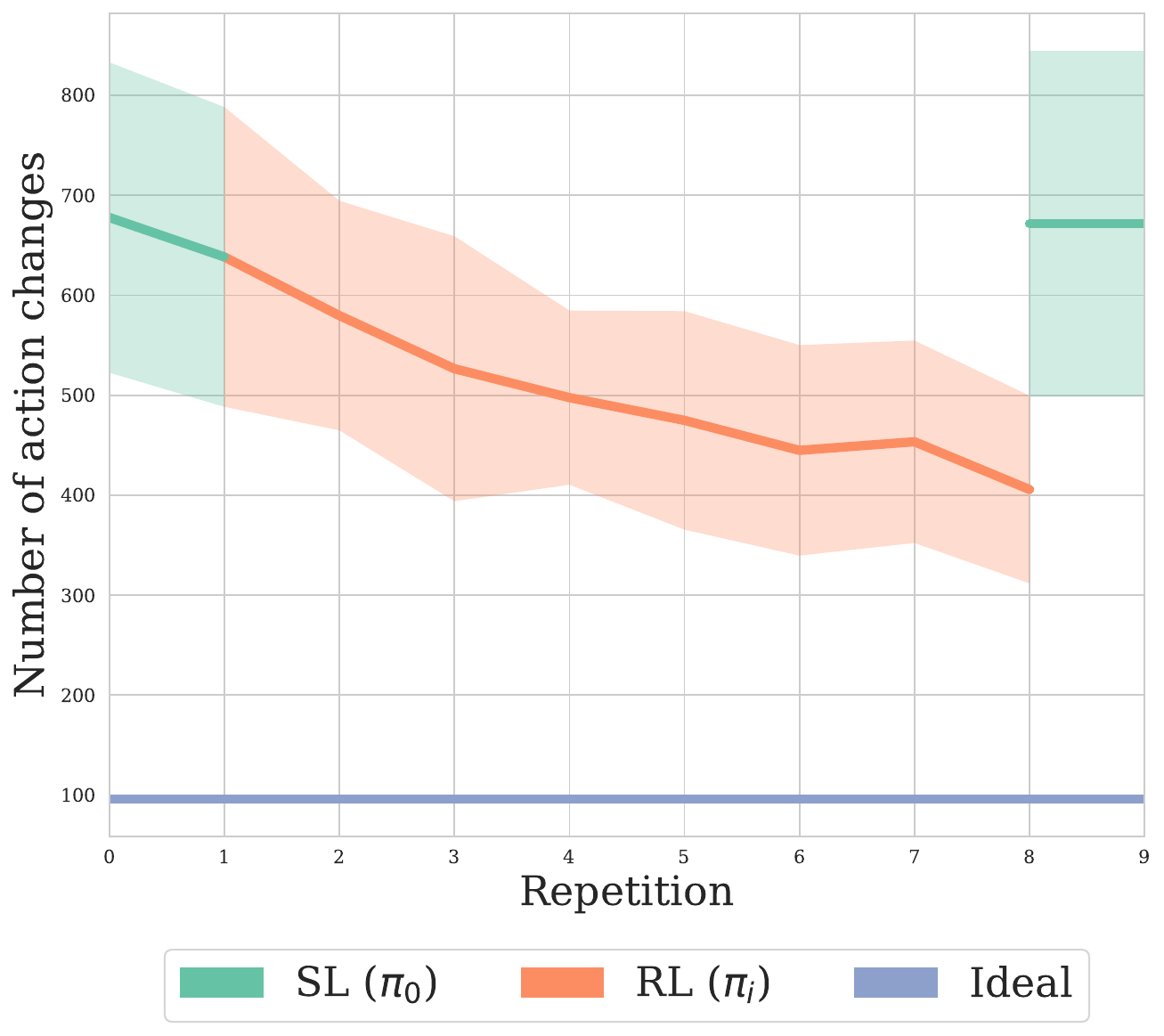}}
\caption{Mean and standard deviation for number of action changes during gameplay over all repetitions. Note that the optimal number of changes for the test song is 96. The number of changes consistently decreases for all participants, even for the lower outliers found in Fig. \ref{fig:rl-cumulative-reward} and Fig. \ref{fig:results-per-dof}. }
\label{fig:rl-num-action-changes}
\end{center}
\end{figure}
\subsection{What does the RL policy learn?}

Upon further inspection of the results, we noticed that the greatest improvement was found in single DOF classes, whereas movements involving 2 or 3 simultaneous DOFs displayed less marked changes (see Fig. \ref{fig:results-per-dof}a-d). We believe that this might be because single DOF movements are more challenging to decode due to their relatively lower muscle activation compared to movements with multiple DOFs, explaining the relatively lower EMR and F1 scores. On one hand, lower scores inherently provide more room for improvement. On the other hand, this could also mean that the RL approach primarily focuses on refining movements that initially perform poorly. These trends are consistent across both the game environment and the \textit{Motion Test}, with slightly more pronounced effects observed in the game environment.

A noticeable outcome of RL training is the reduction in policy prediction changes, indicating improved stability, as shown in Fig. \ref{fig:rl-num-action-changes}. The average changes decrease from 672 with $\pi_0$ to 406 with $\pi_8$, consistently observed across all users, including outliers depicted in Fig. \ref{fig:rl-cumulative-reward} and Fig. \ref{fig:results-per-dof}. This stability induces confidence in the system's performance, making the policy's behavior more predictable for the user.

\begin{figure}[t]
\begin{center}
\centerline{\includegraphics[width=0.7\columnwidth]{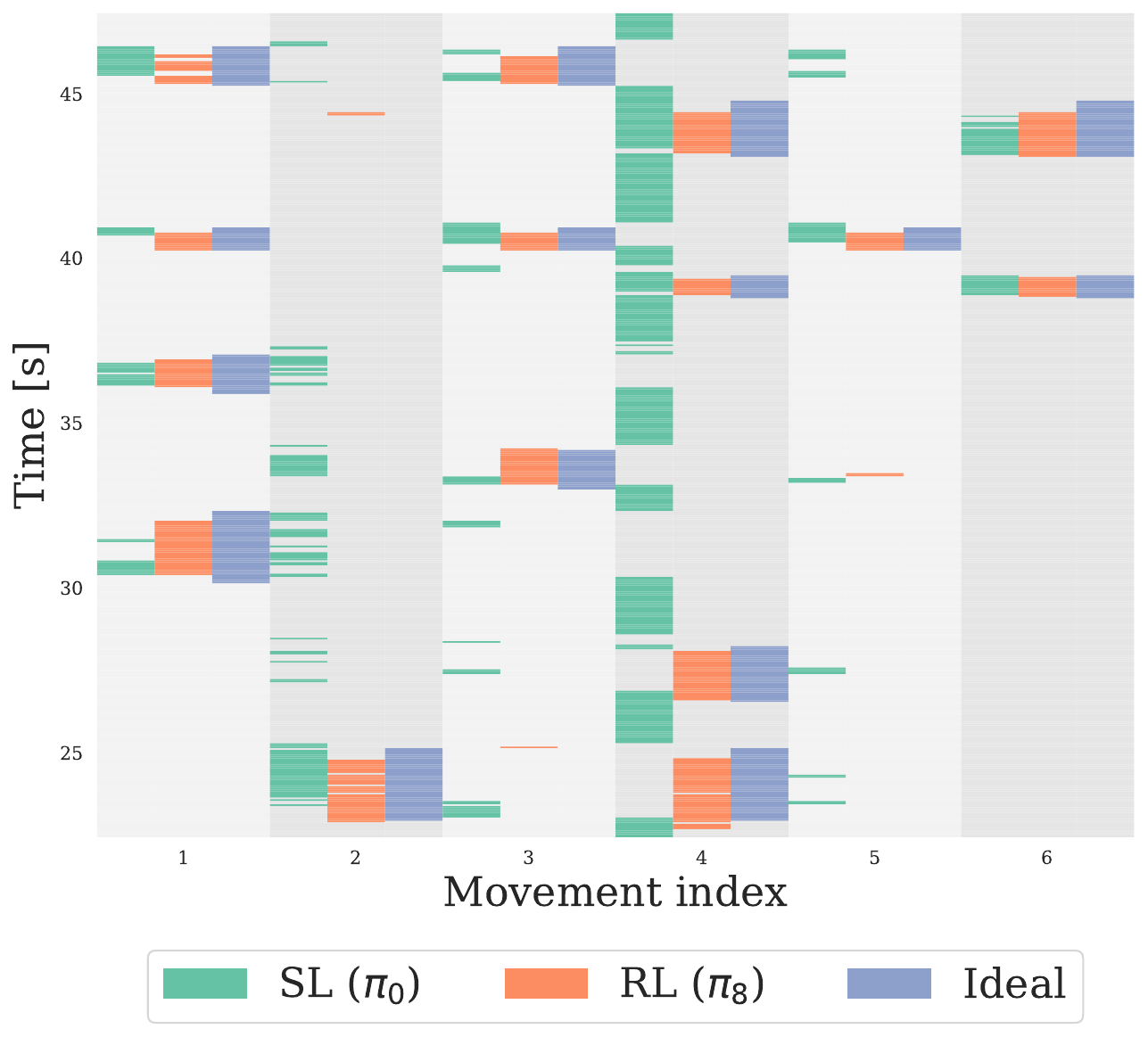}}
\caption{Snapshot of predicted gameplay actions for subject 10 in the last two repetitions of the serious game. As in Fig. \ref{fig:rl-num-action-changes}, green is the pretrained SL policy $\pi_0$, orange after RL training ($\pi_8$) and in grey the ideal predictions. It can be seen that RL predicts actions much more consistently, nearly matching the ideal actions. Subject 10 was chosen as their return increased the most during gameplay, to illustrate the ideal outcome of our method. The time-frame was chosen randomly. }
\label{fig:rl-gameplay-snapshot}
\end{center}
\vspace{-4mm}
\end{figure}

While this trend was observed across all participants, Fig. \ref{fig:rl-gameplay-snapshot} offers an illustrative snapshot chosen at random of predicted actions for subject 10 during gameplay. Actions are predicted more accurately and cohesively with $\pi_8$, as demonstrated by the contiguous blocks of predictions, displaying a marked improvement in stability over the more intermittent predictions associated with $\pi_0$.

\subsection{When is the RL policy unable to learn?}

Although substantial improvements were observed in most cases, there were two low outliers depicted as circles in Fig. \ref{fig:rl-cumulative-reward} and Fig. \ref{fig:results-per-dof}, which we aim to analyze in this section. Participant 3 corresponds to the lowest outlier and actually showed a decrease in performance with the RL policy. Participant 12 corresponds to the second lowest outlier, but still showed a significant improvement using our method. However, since this participant had an under-performing SL policy (lowest across all participants, closely followed by participant 3), even with this improvement, the final RL policy still resulted in a comparatively low performance.

These outliers motivated us to take a closer look at the results and investigate which factors may influence the success of our RL-based procedure. We hypothesize that the poor performance for participants 3 and 12 was due to either \textbf{(a)} low quality EMG data, which can be caused by several factors like electrode placement and noisy connections, or \textbf{(b)} high variability in the participant's muscle activations, including incorrect movements, which could complicate training. 

If hypothesis \textbf{(a)} is true, then there should be an impact throughout the whole experiment, assuming that EMG signal quality did not change during a session. Thus, both pretraining and gameplay data would be affected. This seemed like a valid explanation, since in fact, both participants had an initial policy $\pi_0$ with poor performance, as evidenced by the \textit{Motion Test} EMR of 0.22 for participant 3 and 0.085 for participant 12 (worst two across all participants, see Fig. \ref{fig:entropies}). For participant 3 in particular, performance did not increase during gameplay, which would also be in line with this assumption. However, when looking at the Signal to Noise Ratio (SNR) values it becomes clear that both outliers have similar values to the rest of the participants, as shown in Table \ref{tab:signal_noise_ratios}, indicating that the EMG signal was not the root cause of the poor results. Refer to Appendix \ref{subsec:snr} for details on how the SNR values are calculated for our analysis. 
\begin{table}[hb]
\caption{\textup{SNR for pretraining and mean and standard deviation of SNR for each participant over RL repetitions. For comparison, the \textit{Motion Test} EMRs for $\pi_0$ and $\pi_8$ are shown below. Outliers are highlighted in gray color.}}
\label{tab:signal_noise_ratios}
\begin{center}
\begin{small}
\begin{sc}
\begin{tabular}{ccccc}
\toprule
Person ID & \multicolumn{2}{c}{SL} & \multicolumn{2}{c}{RL} \\
\midrule
 & SNR & EMR & SNR & EMR \\
\midrule
1 & $8.48$ & $0.33$ & $15.28 \pm 1.40$ & $0.58$ \\
2 & $25.44$ & $0.38$ & $10.56 \pm 3.35$ & $0.52$ \\
\rowcolor{Gray}
3 & $17.04$ & $0.22$ & $11.49 \pm 2.67$ & $0.17$ \\
4 & $22.52$ & $0.33$ & $\mathbf{16.94} \pm 3.53$ & $0.66$ \\
5 & $16.54$ & $0.68$ & $13.24 \pm 1.76$ & $0.71$ \\
6 & $10.42$ & $0.45$ & $12.38 \pm 1.24$ & $0.60$ \\
7 & $7.69$ & $0.50$ & $15.74 \pm 1.11$ & $0.74$ \\
8 & $10.16$ & $0.35$ & $14.54 \pm 1.27$ & $0.61$ \\
9 & $17.28$ & $0.31$ & $5.07 \pm 3.82$ & $0.72$ \\
10 & $19.67$ & $0.49$ & $15.52 \pm 1.60$ & $0.66$ \\
11 & $10.46$ & $0.39$ & $12.31 \pm 1.41$ & $0.56$ \\
\rowcolor{Gray}
12 & $\mathbf{31.88}$ & $0.09$ & $11.81 \pm 1.03$ & $0.23$ \\
13 & $9.52$ & $0.64$ & $16.29 \pm 1.92$ & $0.72$ \\
14 & $9.83$ & $0.59$ & $13.71 \pm 2.19$ & $0.72$ \\
15 & $7.22$ & $\mathbf{0.73}$ & $11.46 \pm 2.58$ & $\mathbf{0.83}$ \\

\bottomrule

\end{tabular}
\end{sc}
\end{small}
\end{center}
\vspace{-2mm}
\end{table}

This leads us to investigate the validity of hypothesis \textbf{(b)}, i.e., that the participants' inconsistency, rather than poor signal quality, is the primary factor. To this end, we calculate the Mutual Information (MI) between the recorded features and the ideal labels $\mathbb{I}(s;a^*)$ for each repetition during online training. Notably, we observe that the MI increases across gameplay, with the highest mean MI observed in repetition 8 (See Fig. \ref{fig:mi_gameplay} in the Appendix). This indicates that participants are able to generate signals that are more aligned with their intention over time. However, the MI drops in repetition 9 with $\pi_0$, indicating that the policy influences how much information a user is able to convey about ideal labels, $a^*$.

It is important to highlight that the MI is independent of both the policy and how it is trained, as it only quantifies the amount of information about the ideal labels present in the features. Further details on the MI calculation can be found in Appendix \ref{subsec:mi}. To better understand how MI relates with classification performance, we look at mean gameplay MI values versus EMR in the final repetitions of the experiments.


\begin{figure*}[t]
\begin{center}
\centerline{\includegraphics[width=0.95\linewidth]{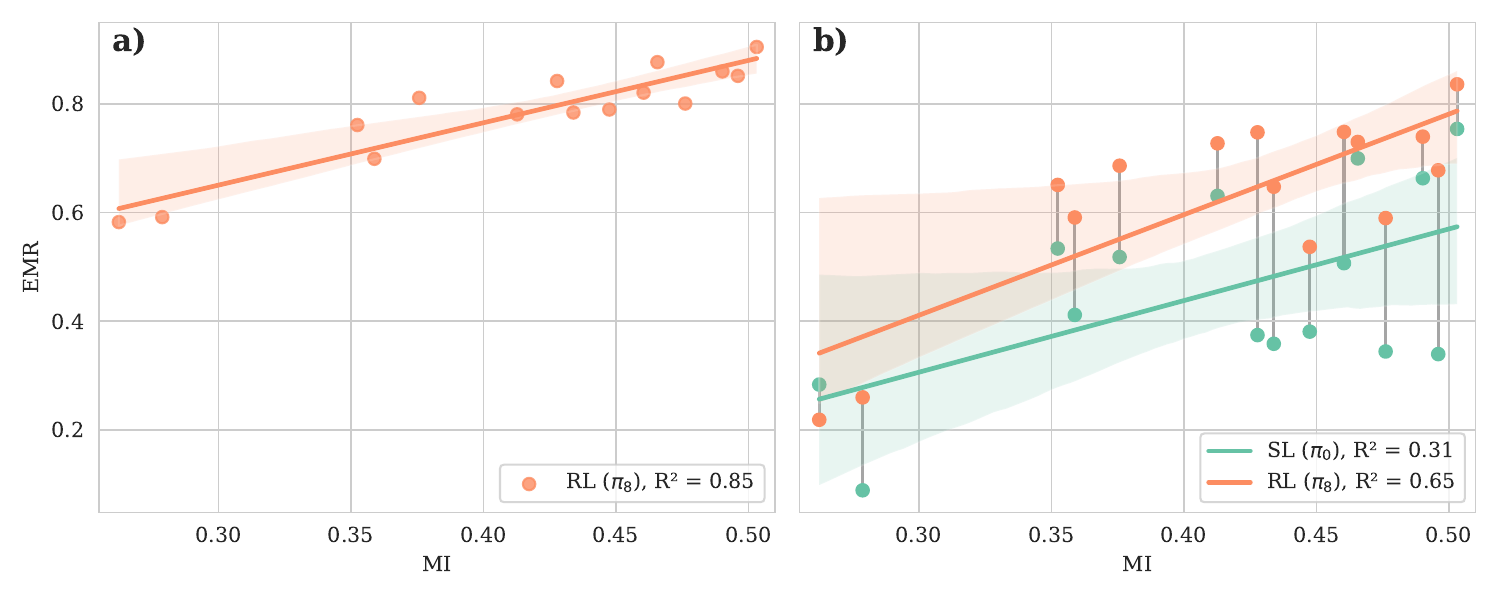}}
\caption{MI values are calculated as the mean over repetition 8, where $\pi_8$ was tested. These are then plotted against the EMR values taken from \textbf{a)} gameplay from repetition 8 and \textbf{b)} the \textit{Motion Test} for both $\pi_8$ and $\pi_0$. Note that for the latter plot, the $x$ and $y$ axes represent separate datasets, one from the gameplay ($x$) and the other from the \textit{Motion Test} ($y$). As both plots indicate, the EMR increases proportionally with the MI during gameplay, suggesting that it can be used as a reasonable predictor of our method's training success. Notably, the two outliers in EMR also exhibit the lowest MI. Furthermore, despite corresponding to different datasets, a relatively similar trend is observed in the \textit{Motion Test} EMR with $\pi_8$, albeit with a slightly worse fit. Conversely, the EMR with $\pi_0$ does not have a strong correlation with the gameplay MI indicating that the SL policy is affected by a distributional shift.} 
\label{fig:mi}
\end{center}
\vspace{-5mm}
\end{figure*}

In Fig. \ref{fig:mi}a), we plot the mean gameplay MI for repetition 8 versus the corresponding gameplay EMR for 
RL policy $\pi_8$ and, fit a linear function to the data. This plot clearly shows that the success of our method increases proportionally with the MI, given the resulting strong linear fit ($R^2=0.85$).

In in Fig. \ref{fig:mi}b), we further compare the mean gameplay MI for repetition 8 with the \textit{Motion Test} EMR for both the SL policy, $\pi_0$, and the RL policy, $\pi_8$. While the gameplay MI also seems to be a reasonable indicator for the \textit{Motion Test} EMR with $\pi_8$ ($R^2=0.65$), this relationship is not as strong with $\pi_0$ ($R^2=0.31$). 
Given that the gameplay MI with $\pi_8$ is predictive of the respective \textit{Motion Test} EMR, a natural question to ask is whether the gameplay MI with $\pi_0$ is similarly predictive of the \textit{Motion Test} EMR of $\pi_0$. We investigate this question using Fig. \ref{fig:mi_sl} in the Appendix and find that the gameplay MI with $\pi_0$ is less predictive of its performance compared to $\pi_8$.

Both plots in Fig. \ref{fig:mi} indicate that indeed hypothesis \textbf{(b)} is the most likely explanation for the two outliers, as they correspond to the lowest MI values. It is worth noting that some human error and inconsistencies occur for all participants during the training process. While this may prolong the learning time, it does not seem to hinder improvement for the majority of participants. Moreover, there may be a threshold in MI somewhere between [0.28, 0.35], below which pattern recognition methods begin to struggle to overcome the lack of consistent information. Importantly, that does not necessarily mean that the participants performed unintended movements, rather that the variability of each movement was higher than for other participants. A promising direction for future work would be to utilize these findings to refine our method.

Finally, in addition to MI we also investigated the Population Stability Index (PSI) in Appendix \ref{subsec:psi}, and explored how variation in contraction force correlates to learning success (see Appendix \ref{subsec:contraction} for details).


\section{Conclusion}\label{sec:conclusion}

In this study, we proposed an RL-based procedure to improve the decoding of motion intent with the aim of creating more intuitive and responsive myoelectric controllers. Our approach using an interactive game significantly enhanced the online, human-in-the-loop performance of an ML controller for all evaluation metrics, thus bridging the gap between offline training and real-life usage.

Since our results are promising, future investigations will be carried out on individuals with amputation, which will provide valuable insights into the effectiveness of our approach in a clinical setting. Additionally, our method exhibits a versatile nature, allowing for its extension to other serious games that more accurately simulate daily prosthetic use, thereby ensuring a higher level of realism and applicability.

Finally, our work opens up avenues for further advancements by incorporating fine-tuning techniques based on human feedback. This has the potential to greatly improve prosthetic functionality as it enables individuals to train on any tasks encountered in their daily lives.


\clearpage
\acknowledgments{This project was funded by the Promobilia Foundation, the IngaBritt and Arne Lundbergs Foundation and the Swedish Research Council (Vetenskapsr{\aa}det). 

Additionally, this work was supported by the Wallenberg AI, Autonomous Systems and Software Program (WASP) funded by the Knut and Alice Wallenberg Foundation.}


\bibliography{main}  

\begin{thebibliography}{47}
\providecommand{\natexlab}[1]{#1}
\providecommand{\url}[1]{\texttt{#1}}
\expandafter\ifx\csname urlstyle\endcsname\relax
  \providecommand{\doi}[1]{doi: #1}\else
  \providecommand{\doi}{doi: \begingroup \urlstyle{rm}\Url}\fi

\bibitem[McDonald et~al.(2021)McDonald, Westcott-McCoy, Weaver, Haagsma, and Kartin]{mcdonald2021global}
C.~L. McDonald, S.~Westcott-McCoy, M.~R. Weaver, J.~Haagsma, and D.~Kartin.
\newblock Global prevalence of traumatic non-fatal limb amputation.
\newblock \emph{Prosthetics and orthotics international}, page 0309364620972258, 2021.

\bibitem[Cordella et~al.(2016)Cordella, Ciancio, Sacchetti, Davalli, Cutti, Guglielmelli, and Zollo]{cordella2016literature}
F.~Cordella, A.~L. Ciancio, R.~Sacchetti, A.~Davalli, A.~G. Cutti, E.~Guglielmelli, and L.~Zollo.
\newblock Literature review on needs of upper limb prosthesis users.
\newblock \emph{Frontiers in neuroscience}, 10:\penalty0 209, 2016.

\bibitem[Smail et~al.(2021)Smail, Neal, Wilkins, and Packham]{smail2021comfort}
L.~C. Smail, C.~Neal, C.~Wilkins, and T.~L. Packham.
\newblock Comfort and function remain key factors in upper limb prosthetic abandonment: findings of a scoping review.
\newblock \emph{Disability and rehabilitation: Assistive technology}, 16\penalty0 (8):\penalty0 821--830, 2021.

\bibitem[Jiang et~al.(2010)Jiang, Falla, d'Avella, Graimann, and Farina]{jiang2010myoelectric}
N.~Jiang, D.~Falla, A.~d'Avella, B.~Graimann, and D.~Farina.
\newblock Myoelectric control in neurorehabilitation.
\newblock \emph{Critical Reviews™ in Biomedical Engineering}, 38\penalty0 (4), 2010.

\bibitem[Li et~al.(2021)Li, Shi, and Yu]{li2021gesture}
W.~Li, P.~Shi, and H.~Yu.
\newblock Gesture recognition using surface electromyography and deep learning for prostheses hand: state-of-the-art, challenges, and future.
\newblock \emph{Frontiers in neuroscience}, 15:\penalty0 621885, 2021.

\bibitem[Kuiken et~al.(2016)Kuiken, Miller, Turner, and Hargrove]{kuiken2016comparison}
T.~A. Kuiken, L.~A. Miller, K.~Turner, and L.~J. Hargrove.
\newblock A comparison of pattern recognition control and direct control of a multiple degree-of-freedom transradial prosthesis.
\newblock \emph{IEEE journal of translational engineering in health and medicine}, 4:\penalty0 1--8, 2016.

\bibitem[Mereu et~al.(2021)Mereu, Leone, Gentile, Cordella, Gruppioni, and Zollo]{mereu2021control}
F.~Mereu, F.~Leone, C.~Gentile, F.~Cordella, E.~Gruppioni, and L.~Zollo.
\newblock Control strategies and performance assessment of upper-limb tmr prostheses: a review.
\newblock \emph{Sensors}, 21\penalty0 (6):\penalty0 1953, 2021.

\bibitem[Oskoei and Hu(2008)]{oskoei2008support}
M.~A. Oskoei and H.~Hu.
\newblock Support vector machine-based classification scheme for myoelectric control applied to upper limb.
\newblock \emph{IEEE transactions on biomedical engineering}, 55\penalty0 (8):\penalty0 1956--1965, 2008.

\bibitem[Hudgins et~al.(1993)Hudgins, Parker, and Scott]{hudgins1993new}
B.~Hudgins, P.~Parker, and R.~N. Scott.
\newblock A new strategy for multifunction myoelectric control.
\newblock \emph{IEEE transactions on biomedical engineering}, 40\penalty0 (1):\penalty0 82--94, 1993.

\bibitem[Williams et~al.(2022)Williams, Shehata, Dawson, Scheme, Hebert, and Pilarski]{williams2022recurrent}
H.~E. Williams, A.~W. Shehata, M.~R. Dawson, E.~Scheme, J.~S. Hebert, and P.~M. Pilarski.
\newblock Recurrent convolutional neural networks as an approach to position-aware myoelectric prosthesis control.
\newblock \emph{IEEE Transactions on Biomedical Engineering}, 69\penalty0 (7):\penalty0 2243--2255, 2022.

\bibitem[Luu et~al.(2022)Luu, Nguyen, Jiang, Drealan, Xu, Wu, Tam, Zhao, Lim, Overstreet, et~al.]{luu2022artificial}
D.~K. Luu, A.~T. Nguyen, M.~Jiang, M.~W. Drealan, J.~Xu, T.~Wu, W.-k. Tam, W.~Zhao, B.~Z. Lim, C.~K. Overstreet, et~al.
\newblock Artificial intelligence enables real-time and intuitive control of prostheses via nerve interface.
\newblock \emph{IEEE Transactions on Biomedical Engineering}, 69\penalty0 (10):\penalty0 3051--3063, 2022.

\bibitem[Cha et~al.(2022)Cha, An, Choi, Yang, Park, and Park]{cha2022study}
H.~Cha, S.~An, S.~Choi, S.~Yang, S.~Park, and S.~Park.
\newblock Study on intention recognition and sensory feedback: Control of robotic prosthetic hand through emg classification and proprioceptive feedback using rule-based haptic device.
\newblock \emph{IEEE Transactions on Haptics}, 15\penalty0 (3):\penalty0 560--571, 2022.

\bibitem[Godoy et~al.(2022)Godoy, Dwivedi, and Liarokapis]{godoy2022electromyography}
R.~V. Godoy, A.~Dwivedi, and M.~Liarokapis.
\newblock Electromyography based decoding of dexterous, in-hand manipulation motions with temporal multichannel vision transformers.
\newblock \emph{IEEE Transactions on Neural Systems and Rehabilitation Engineering}, 30:\penalty0 2207--2216, 2022.

\bibitem[Ortiz-Catalan et~al.(2015)Ortiz-Catalan, Rouhani, Br{\aa}nemark, and H{\aa}kansson]{ortiz2015offline}
M.~Ortiz-Catalan, F.~Rouhani, R.~Br{\aa}nemark, and B.~H{\aa}kansson.
\newblock Offline accuracy: a potentially misleading metric in myoelectric pattern recognition for prosthetic control.
\newblock In \emph{2015 37th Annual International Conference of the IEEE Engineering in Medicine and Biology Society (EMBC)}, pages 1140--1143. IEEE, 2015.

\bibitem[Mouchoux et~al.(2021)Mouchoux, Carisi, Dosen, Farina, Schilling, and Markovic]{mouchoux2021artificial}
J.~Mouchoux, S.~Carisi, S.~Dosen, D.~Farina, A.~F. Schilling, and M.~Markovic.
\newblock Artificial perception and semiautonomous control in myoelectric hand prostheses increases performance and decreases effort.
\newblock \emph{IEEE Transactions on Robotics}, 37\penalty0 (4):\penalty0 1298--1312, 2021.

\bibitem[Vaskov et~al.(2022)Vaskov, Vu, North, Davis, Kung, Gates, Cederna, and Chestek]{vaskov2022surgically}
A.~K. Vaskov, P.~P. Vu, N.~North, A.~J. Davis, T.~A. Kung, D.~H. Gates, P.~S. Cederna, and C.~A. Chestek.
\newblock Surgically implanted electrodes enable real-time finger and grasp pattern recognition for prosthetic hands.
\newblock \emph{IEEE Transactions on Robotics}, 38\penalty0 (5):\penalty0 2841--2857, 2022.

\bibitem[Tommasi et~al.(2012)Tommasi, Orabona, Castellini, and Caputo]{tommasi2012improving}
T.~Tommasi, F.~Orabona, C.~Castellini, and B.~Caputo.
\newblock Improving control of dexterous hand prostheses using adaptive learning.
\newblock \emph{IEEE Transactions on Robotics}, 29\penalty0 (1):\penalty0 207--219, 2012.

\bibitem[BAKIRCIO{\u{G}}LU and {\"O}zkurt(2020)]{bakirciouglu2020classification}
K.~BAKIRCIO{\u{G}}LU and N.~{\"O}zkurt.
\newblock Classification of emg signals using convolution neural network.
\newblock \emph{International Journal of Applied Mathematics Electronics and Computers}, 8\penalty0 (4):\penalty0 115--119, 2020.

\bibitem[Zbinden et~al.(2024)Zbinden, Molin, and Ortiz-Catalan]{zbinden2024deep}
J.~Zbinden, J.~Molin, and M.~Ortiz-Catalan.
\newblock Deep learning for enhanced prosthetic control: Real-time motor intent decoding for simultaneous control of artificial limbs.
\newblock \emph{IEEE Transactions on Neural Systems and Rehabilitation Engineering}, 2024.

\bibitem[labs~at Reality~Labs et~al.(2024)labs~at Reality~Labs, Sussillo, Kaifosh, and Reardon]{ctrl2024generic}
C.~labs~at Reality~Labs, D.~Sussillo, P.~Kaifosh, and T.~Reardon.
\newblock A generic noninvasive neuromotor interface for human-computer interaction.
\newblock \emph{bioRxiv}, pages 2024--02, 2024.

\bibitem[Ross et~al.(2011)Ross, Gordon, and Bagnell]{ross2011reduction}
S.~Ross, G.~Gordon, and D.~Bagnell.
\newblock A reduction of imitation learning and structured prediction to no-regret online learning.
\newblock In \emph{Proceedings of the fourteenth international conference on artificial intelligence and statistics}, pages 627--635. JMLR Workshop and Conference Proceedings, 2011.

\bibitem[Gijsberts et~al.(2014)Gijsberts, Bohra, Sierra~Gonz{\'a}lez, Werner, Nowak, Caputo, Roa, and Castellini]{gijsberts2014stable}
A.~Gijsberts, R.~Bohra, D.~Sierra~Gonz{\'a}lez, A.~Werner, M.~Nowak, B.~Caputo, M.~A. Roa, and C.~Castellini.
\newblock Stable myoelectric control of a hand prosthesis using non-linear incremental learning.
\newblock \emph{Frontiers in neurorobotics}, 8:\penalty0 8, 2014.

\bibitem[Ketyk{\'o} et~al.(2019)Ketyk{\'o}, Kov{\'a}cs, and Varga]{ketyko2019domain}
I.~Ketyk{\'o}, F.~Kov{\'a}cs, and K.~Z. Varga.
\newblock Domain adaptation for semg-based gesture recognition with recurrent neural networks.
\newblock In \emph{2019 International Joint Conference on Neural Networks (IJCNN)}, pages 1--7. IEEE, 2019.

\bibitem[Chen et~al.(2020)Chen, Li, Hu, Zhang, and Chen]{chen2020hand}
X.~Chen, Y.~Li, R.~Hu, X.~Zhang, and X.~Chen.
\newblock Hand gesture recognition based on surface electromyography using convolutional neural network with transfer learning method.
\newblock \emph{IEEE Journal of Biomedical and Health Informatics}, 25\penalty0 (4):\penalty0 1292--1304, 2020.

\bibitem[Du et~al.(2017)Du, Jin, Wei, Hu, and Geng]{du2017surface}
Y.~Du, W.~Jin, W.~Wei, Y.~Hu, and W.~Geng.
\newblock Surface emg-based inter-session gesture recognition enhanced by deep domain adaptation.
\newblock \emph{Sensors}, 17\penalty0 (3):\penalty0 458, 2017.

\bibitem[Cote-Allard et~al.(2021)Cote-Allard, Gagnon-Turcotte, Phinyomark, Glette, Scheme, Laviolette, and Gosselin]{cote2021transferable}
U.~Cote-Allard, G.~Gagnon-Turcotte, A.~Phinyomark, K.~Glette, E.~Scheme, F.~Laviolette, and B.~Gosselin.
\newblock A transferable adaptive domain adversarial neural network for virtual reality augmented emg-based gesture recognition.
\newblock \emph{IEEE Transactions on Neural Systems and Rehabilitation Engineering}, 29:\penalty0 546--555, 2021.

\bibitem[Pilarski et~al.(2011)Pilarski, Dawson, Degris, Fahimi, Carey, and Sutton]{pilarski2011online}
P.~M. Pilarski, M.~R. Dawson, T.~Degris, F.~Fahimi, J.~P. Carey, and R.~S. Sutton.
\newblock Online human training of a myoelectric prosthesis controller via actor-critic reinforcement learning.
\newblock In \emph{2011 IEEE international conference on rehabilitation robotics}, pages 1--7. IEEE, 2011.

\bibitem[Vasan and Pilarski(2017)]{vasan2017learning}
G.~Vasan and P.~M. Pilarski.
\newblock Learning from demonstration: Teaching a myoelectric prosthesis with an intact limb via reinforcement learning.
\newblock In \emph{2017 International Conference on Rehabilitation Robotics (ICORR)}, pages 1457--1464. IEEE, 2017.

\bibitem[Rosenbaum(2009)]{rosenbaum2009human}
D.~A. Rosenbaum.
\newblock \emph{Human motor control}.
\newblock Academic press, 2nd edition, 2009.

\bibitem[Osborn et~al.(2021)Osborn, Moran, Johannes, Sutton, Wormley, Dohopolski, Nordstrom, Butkus, Chi, Pasquina, et~al.]{osborn2021extended}
L.~E. Osborn, C.~W. Moran, M.~S. Johannes, E.~E. Sutton, J.~M. Wormley, C.~Dohopolski, M.~J. Nordstrom, J.~A. Butkus, A.~Chi, P.~F. Pasquina, et~al.
\newblock Extended home use of an advanced osseointegrated prosthetic arm improves function, performance, and control efficiency.
\newblock \emph{Journal of neural engineering}, 18\penalty0 (2):\penalty0 026020, 2021.

\bibitem[Zbinden et~al.(2023)Zbinden, Sassu, Mastinu, Eric~J., Munoz-Novoa, Br{\aa}nemark, and Ortiz-Catalan]{zbinden2023a}
J.~Zbinden, P.~Sassu, E.~Mastinu, E.~Eric~J., M.~Munoz-Novoa, R.~Br{\aa}nemark, and M.~Ortiz-Catalan.
\newblock Improved control of a prosthetic limb by surgically creating electro-neuromuscular constructs with implanted electrodes.
\newblock \emph{Science Translational Medicine}, 15\penalty0 (704):\penalty0 eabq3665, 2023.

\bibitem[Zia~ur Rehman et~al.(2018)Zia~ur Rehman, Waris, Gilani, Jochumsen, Niazi, Jamil, Farina, and Kamavuako]{zia2018multiday}
M.~Zia~ur Rehman, A.~Waris, S.~O. Gilani, M.~Jochumsen, I.~K. Niazi, M.~Jamil, D.~Farina, and E.~N. Kamavuako.
\newblock Multiday emg-based classification of hand motions with deep learning techniques.
\newblock \emph{Sensors}, 18\penalty0 (8):\penalty0 2497, 2018.

\bibitem[Jiang et~al.(2008)Jiang, Englehart, and Parker]{jiang2008extracting}
N.~Jiang, K.~B. Englehart, and P.~A. Parker.
\newblock Extracting simultaneous and proportional neural control information for multiple-dof prostheses from the surface electromyographic signal.
\newblock \emph{IEEE transactions on Biomedical Engineering}, 56\penalty0 (4):\penalty0 1070--1080, 2008.

\bibitem[Ameri et~al.(2018)Ameri, Akhaee, Scheme, and Englehart]{ameri2018real}
A.~Ameri, M.~A. Akhaee, E.~Scheme, and K.~Englehart.
\newblock Real-time, simultaneous myoelectric control using a convolutional neural network.
\newblock \emph{PloS one}, 13\penalty0 (9):\penalty0 e0203835, 2018.

\bibitem[Levine et~al.(2020)Levine, Kumar, Tucker, and Fu]{levine2020offline}
S.~Levine, A.~Kumar, G.~Tucker, and J.~Fu.
\newblock Offline reinforcement learning: Tutorial, review, and perspectives on open problems.
\newblock \emph{arXiv preprint arXiv:2005.01643}, 2020.

\bibitem[Nair et~al.(2020)Nair, Gupta, Dalal, and Levine]{nair2020awac}
A.~Nair, A.~Gupta, M.~Dalal, and S.~Levine.
\newblock Awac: Accelerating online reinforcement learning with offline datasets.
\newblock \emph{arXiv preprint arXiv:2006.09359}, 2020.

\bibitem[Prahm et~al.(2018)Prahm, Kayali, Sturma, and Aszmann]{prahm2018playbionic}
C.~Prahm, F.~Kayali, A.~Sturma, and O.~Aszmann.
\newblock Playbionic: game-based interventions to encourage patient engagement and performance in prosthetic motor rehabilitation.
\newblock \emph{PM\&R}, 10\penalty0 (11):\penalty0 1252--1260, 2018.

\bibitem[Azam et~al.(2021)Azam, Munir, Rafique, Sheri, Hussain, and Jeon]{azam2021n}
S.~Azam, F.~Munir, M.~A. Rafique, A.~M. Sheri, M.~I. Hussain, and M.~Jeon.
\newblock N 2 c: neural network controller design using behavioral cloning.
\newblock \emph{IEEE Transactions on Intelligent Transportation Systems}, 22\penalty0 (7):\penalty0 4744--4756, 2021.

\bibitem[Kuiken et~al.(2009)Kuiken, Li, Lock, Lipschutz, Miller, Stubblefield, and Englehart]{kuiken2009targeted}
T.~A. Kuiken, G.~Li, B.~A. Lock, R.~D. Lipschutz, L.~A. Miller, K.~A. Stubblefield, and K.~B. Englehart.
\newblock Targeted muscle reinnervation for real-time myoelectric control of multifunction artificial arms.
\newblock \emph{Jama}, 301\penalty0 (6):\penalty0 619--628, 2009.

\bibitem[Yang(1999)]{yang1999evaluation}
Y.~Yang.
\newblock An evaluation of statistical approaches to text categorization.
\newblock \emph{Information retrieval}, 1\penalty0 (1-2):\penalty0 69--90, 1999.

\bibitem[Kumar et~al.(2022)Kumar, Singh, Tian, Finn, and Levine]{kumar2022workflow}
A.~Kumar, A.~Singh, S.~Tian, C.~Finn, and S.~Levine.
\newblock A workflow for offline model-free robotic reinforcement learning.
\newblock In \emph{Conference on Robot Learning}, pages 417--428. PMLR, 2022.

\bibitem[Wilcoxon(1992)]{wilcoxon1992individual}
F.~Wilcoxon.
\newblock \emph{Individual comparisons by ranking methods}.
\newblock Springer, 1992.

\bibitem[Hannius et~al.(2024)Hannius, Laezza, and Zbinden]{hannius2024towards}
A.~Hannius, R.~Laezza, and J.~Zbinden.
\newblock Towards pose invariant bionic limb control: A comparative study of two unsupervised domain adaptation methods.
\newblock \emph{TechRxiv preprint 10.36227/techrxiv.173202841.15026283/v1}, 2024.

\bibitem[Kraskov et~al.(2004)Kraskov, St{\"o}gbauer, and Grassberger]{kraskov2004estimating}
A.~Kraskov, H.~St{\"o}gbauer, and P.~Grassberger.
\newblock Estimating mutual information.
\newblock \emph{Physical Review E—Statistical, Nonlinear, and Soft Matter Physics}, 69\penalty0 (6):\penalty0 066138, 2004.

\bibitem[Ross(2014)]{ross2014mutual}
B.~C. Ross.
\newblock Mutual information between discrete and continuous data sets.
\newblock \emph{PloS one}, 9\penalty0 (2):\penalty0 e87357, 2014.

\bibitem[Pedregosa et~al.(2011)Pedregosa, Varoquaux, Gramfort, Michel, Thirion, Grisel, Blondel, Prettenhofer, Weiss, Dubourg, Vanderplas, Passos, Cournapeau, Brucher, Perrot, and Duchesnay]{scikitlearn2011}
F.~Pedregosa, G.~Varoquaux, A.~Gramfort, V.~Michel, B.~Thirion, O.~Grisel, M.~Blondel, P.~Prettenhofer, R.~Weiss, V.~Dubourg, J.~Vanderplas, A.~Passos, D.~Cournapeau, M.~Brucher, M.~Perrot, and E.~Duchesnay.
\newblock Scikit-learn: Machine learning in {P}ython.
\newblock \emph{Journal of Machine Learning Research}, 12:\penalty0 2825--2830, 2011.

\bibitem[Yurdakul(2018)]{yurdakul2018statistical}
B.~Yurdakul.
\newblock \emph{Statistical properties of population stability index}.
\newblock Western Michigan University, 2018.

\end{thebibliography}

\appendix

\section{Hyperparameter Selection}\label{sec:hyperparameter}
We employed a hyperparameter optimization process to find the most suitable parameters for RL training. This required offline evaluation of policies in order to be done at scale. We simulated playing a song by re-predicting actions of a recorded session with the current policy. This simulation provided the cumulative reward which served as such an offline measure. The policy with the highest simulated return on test data was chosen. We additionally employed the model selection procedure, as described in Section\ref{subsubsec:model_selection}, during the search to chose the best model of each individual trial using its training data. While using training data for model selection is not ideal, it is closer to deployment as there might be a lack of access to a test song. See Table \ref{tab:awac-hyperparameter} for all parameters. Note that we could search for values of $\lambda$ this way, since the AWAC algorithm treats the Lagrange multiplier as a hyperparameter. The results of this hyperparameter search indicate that a high randomization of wrong notes, $\mathcal{\epsilon}$, leads to better offline performance. However, to ensure that is indeed the case, this should also be validated with online tests.

\begin{table}[thb]
\caption{\textup{Hyperparameters used for RL training. The values are found by a hyperparameter search where the policy with highest test return was chosen.}}
\label{tab:awac-hyperparameter}
\begin{center}
\begin{small}
\begin{tabular}{p{6.2cm}p{1.8cm}}
\toprule
\textsc{Hyperparameter} & \textsc{Value} \\
\midrule
Batch size & $512$ \\
Discount factor, $\gamma$ & $0.8935$ \\
\textbf{Lagrange multiplier}, $\lambda$ & \textbf{0.95} \\
Reward scaling & $1$ \\
Policy weight decay & $10^{-4}$ \\
Policy learning rate & $9.844\times10^{-4}$ \\
Q-function hidden sizes  & $[256, 256]$ \\
Q-function hidden activation function & ReLU \\
Q-function weight decay & $0$ \\
Q-function learning rate & $7.627\times10^{-4}$ \\
Number of critics & $2$ \\
Target network synchronization coefficient, $\tau$ & $8.948\times10^{-3}$ \\
Number of actions sampled to calculate $A(s,a)$ & $1$ \\
N-step TD calculation & $1$ \\
Interval to update policy & $4$ \\
\textbf{Chance to randomize wrong notes}, $\mathcal{\epsilon}$ & \textbf{0.9} \\
\bottomrule 
\end{tabular}
\end{small}
\end{center}
\vskip -0.2in
\end{table}

\section{Additional Metrics and Results}
While we presented the most important evaluation metrics in the Section \ref{sec:evaluation}, there were some additional metrics which were used to analyse the results in greater depth. These metrics are presented in the following sections.

\subsection{Signal to Noise Ratio}\label{subsec:snr}
We calculate the SNR as:
\begin{equation}
    \text{SNR}_{k, m_i} = \max_{l} \left[10 \cdot \log_{10}\left(\frac{\text{MAV}_{m_i}}{\text{MAV}_{m_0}}\right) \right]
\end{equation}
where $k$ is the Person ID, and $l$ is the EMG channel index. We take the maximum SNR over all channels due to significant variations in channel activation across movements. MAV serves as a signal strength indicator. The SNR per individual is computed as the maximum across all movements, illustrated in Fig. \ref{fig:movements}. Hence, this measure demonstrates the maximum attainable signal strength.

\subsection{Mutual Information}\label{subsec:mi}
Mutual Information (MI) is a measure of how much information is present in one random variable $X$ about another random variable $Y$. It can be defined as
\begin{align*}
    \mathbb{I}(X;Y) = \mathbb{H}(X) + \mathbb{H}(Y) - \mathbb{H}(X,Y)
\end{align*}
where $\mathbb{H}$ is Shannon's entropy. The MI equates to zero if the two variables are independent and grows in value as the dependence between $X$ and $Y$ increases. To compute the MI, we used a non-parametric method based on entropy estimation from k-nearest neighbors distances, proposed by \cite{kraskov2004estimating, ross2014mutual}. Specifically, we use the \texttt{scikit-learn} implementation \cite{scikitlearn2011} (i.e. \texttt{mutual\_info\_classif}).

\begin{figure}[h]
\begin{center}
\centerline{\includegraphics[width=0.985\columnwidth]{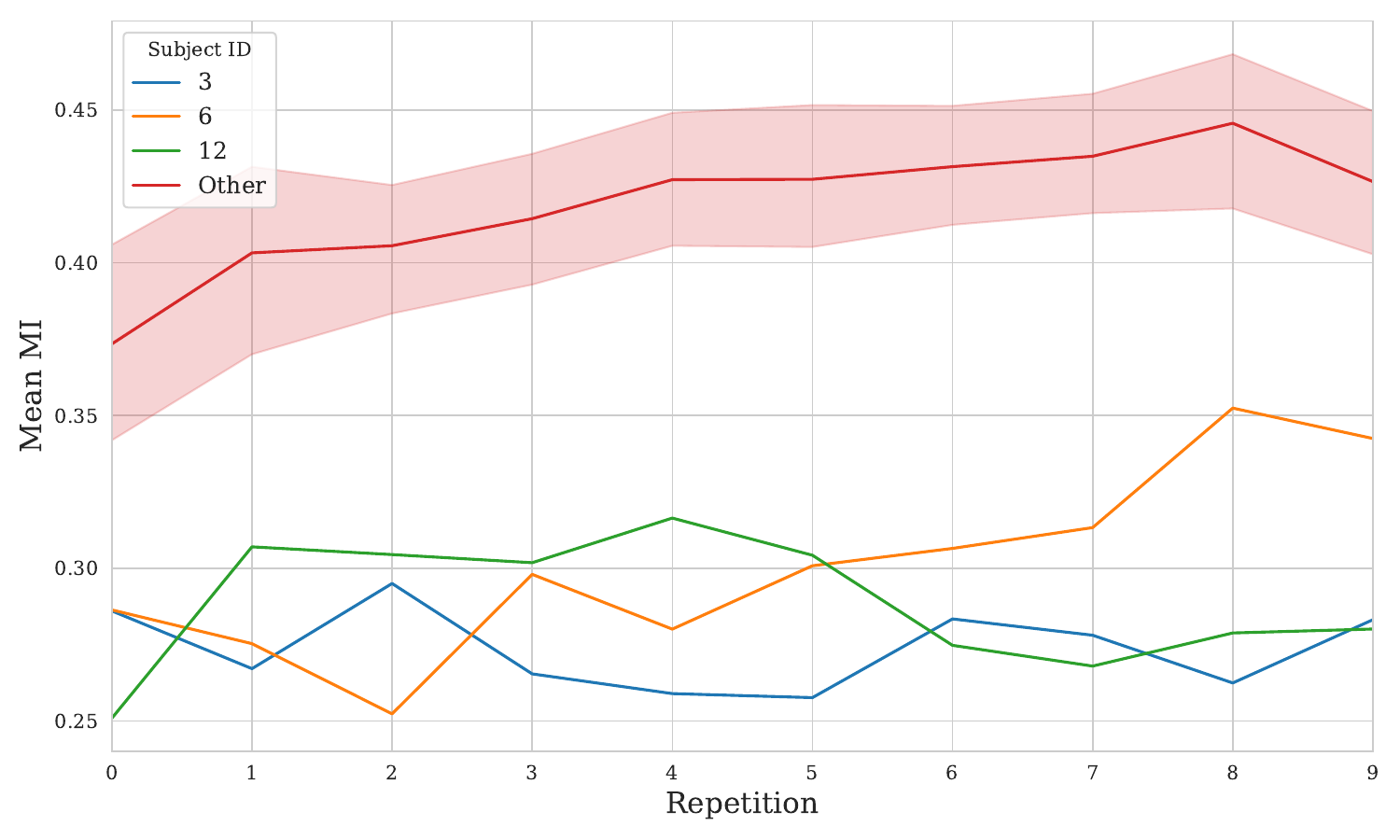}}
\caption{Mean MI $\mathbb{I}(s;a^*)$ over gameplay repetitions. It can be seen that it increases up to repetition 8. The drop in repetition 9 is causes by the change of policy, as here $\pi_0$ is used. This indicates that the quality of a policy $\pi$ impacts the amount of information a user is able to convey about their intention, and that the improvement in MI might not solely be caused by human learning. There are 3 participants with relatively low MI, but different to participant 3 and 12, the MI for participant 6 increased over training. This could be an indicator for why the final EMR for participant 6 was higher than for 3 and 12.} 
\label{fig:mi_gameplay}
\end{center}
\vspace{-5mm}
\end{figure}

For our analysis, we focus on the MI between features and ideal labels, denoted as $\mathbb{I}(s;a^*)$. We first investigate how this value evolves over gameplay repetitions. To this end, we plot the mean MI against the number of repetitions in Fig. \ref{fig:mi_gameplay}. As shown, the MI increases up to repetition 8 and then decreases in repetition 9. Notably, repetition 9 is conducted using policy $\pi_0$, suggesting a potential connection between policy and MI.

\begin{figure}[h]
\begin{center}
\centerline{\includegraphics[width=0.8\columnwidth]{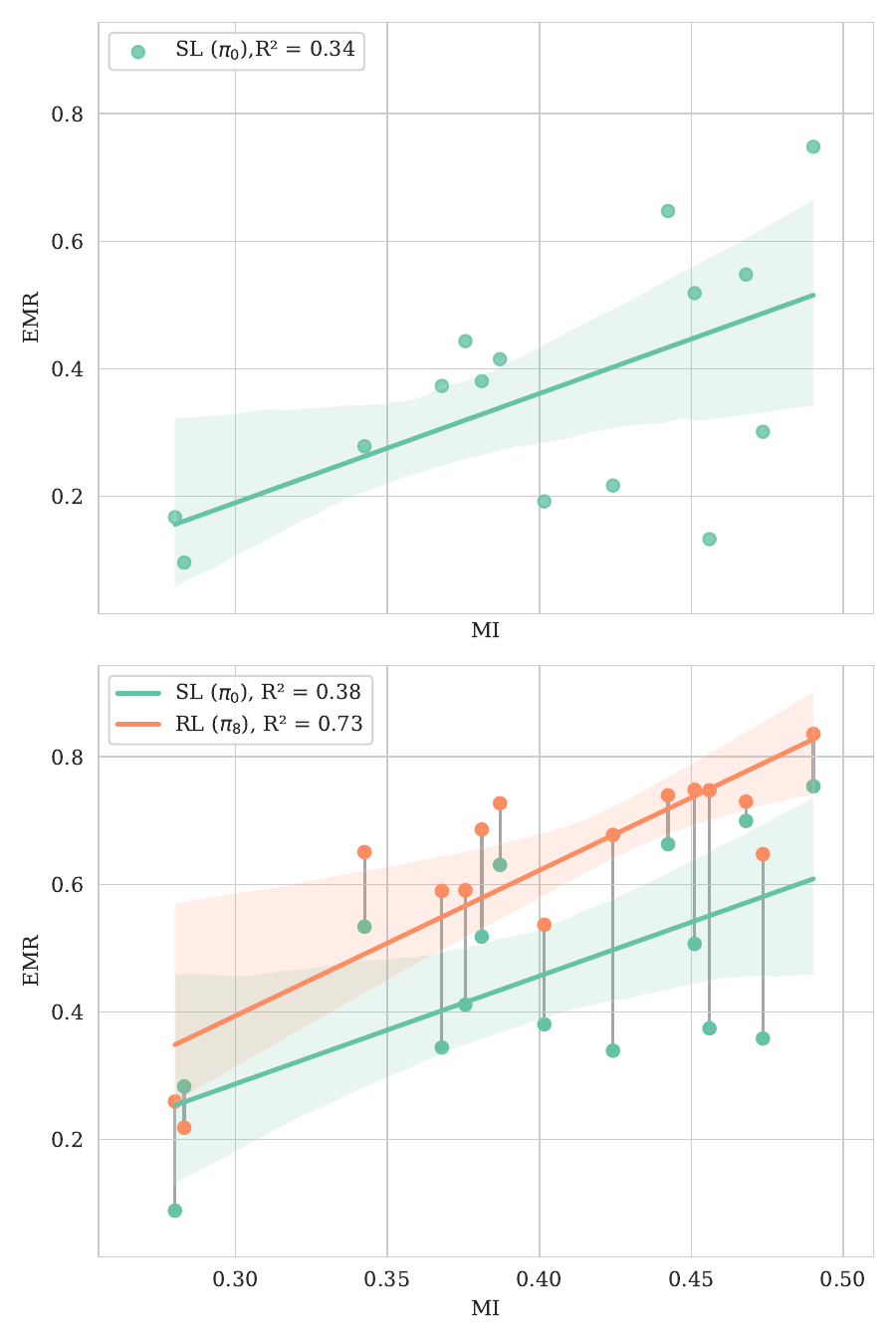}}
\caption{Mean MI $\mathbb{I}(s;a^*)$ for gameplay with $\pi_0$ after training with our method. While there are some trends that could indicate a correlation between MI and EMR in this setting, it less clear than in the same comparison for the MI calculated with $\pi_8$.} 
\label{fig:mi_sl}
\end{center}
\vspace{-5mm}
\end{figure}

Given the observed correlation between the MI in repetition 8 and the success of our method (i.e., the MI in repetition 8 appears predictive of the performance of $\pi_8$), we wanted to investigate whether the MI in repetition 9 (using $\pi_0$) is similarly predictive of the performance of $\pi_0$. To address this, we present a comparable plot for repetition 9 in Fig. \ref{fig:mi_sl}. Our results indicate that the gameplay MI using $\pi_0$ is less predictive of its performance compared to $\pi_8$.

\subsection{Population Stability Index}\label{subsec:psi}
To investigate if inconsistencies in movements where the cause for unsuccessful training in two cases, we also calculate the Population Stability Index (PSI) \cite{yurdakul2018statistical} to further examine how feature distributions drift across each song repetition, for each participant. We compute the PSI of each feature as:
\begin{equation}
    \text{PSI}(P, Q) = \sum_j^{10} (P_j - Q_j) \cdot \ln\left(\frac{P_j}{Q_j}\right)
\end{equation}
where $P$ is the base distribution, $Q$ is the target distribution, and $j$ indicates the bin index. To calculate the PSI, we discretize each feature's distribution into 8 bins using a reference distribution $P$. We use median-based binning, starting with an initial split based on the median, followed by further cuts at intervals of ${1}/{3}$ of the standard deviation, resulting in 4 bins on each side. Additionally, we include a lowest and highest bin with cutoffs at the minimum and maximum of $P$ respectively. Empty edge bins are removed, leaving a final count of 8-10 bins.
As the PSI is defined for individual features, we take the mean over the PSI values to quantify the extent of distribution divergence. We carry out a first test by computing the mean over all participants of $\text{PSI}(\mathcal{S}_0, \mathcal{S}_1)$ and $\text{PSI}(\mathcal{S}_7, \mathcal{S}_8)$, where $\mathcal{S}_i$ indicates the state distribution of repetition $i$. It is noticeable that the PSI between the last two RL repetitions decreased by 59\% (from 0.33 to 0.19) when compared with the first two repetitions. We believe this is due to the increasingly consistent motions of the user. 

\begin{figure}[h]
\begin{center}
\centerline{\includegraphics[width=0.7\columnwidth]{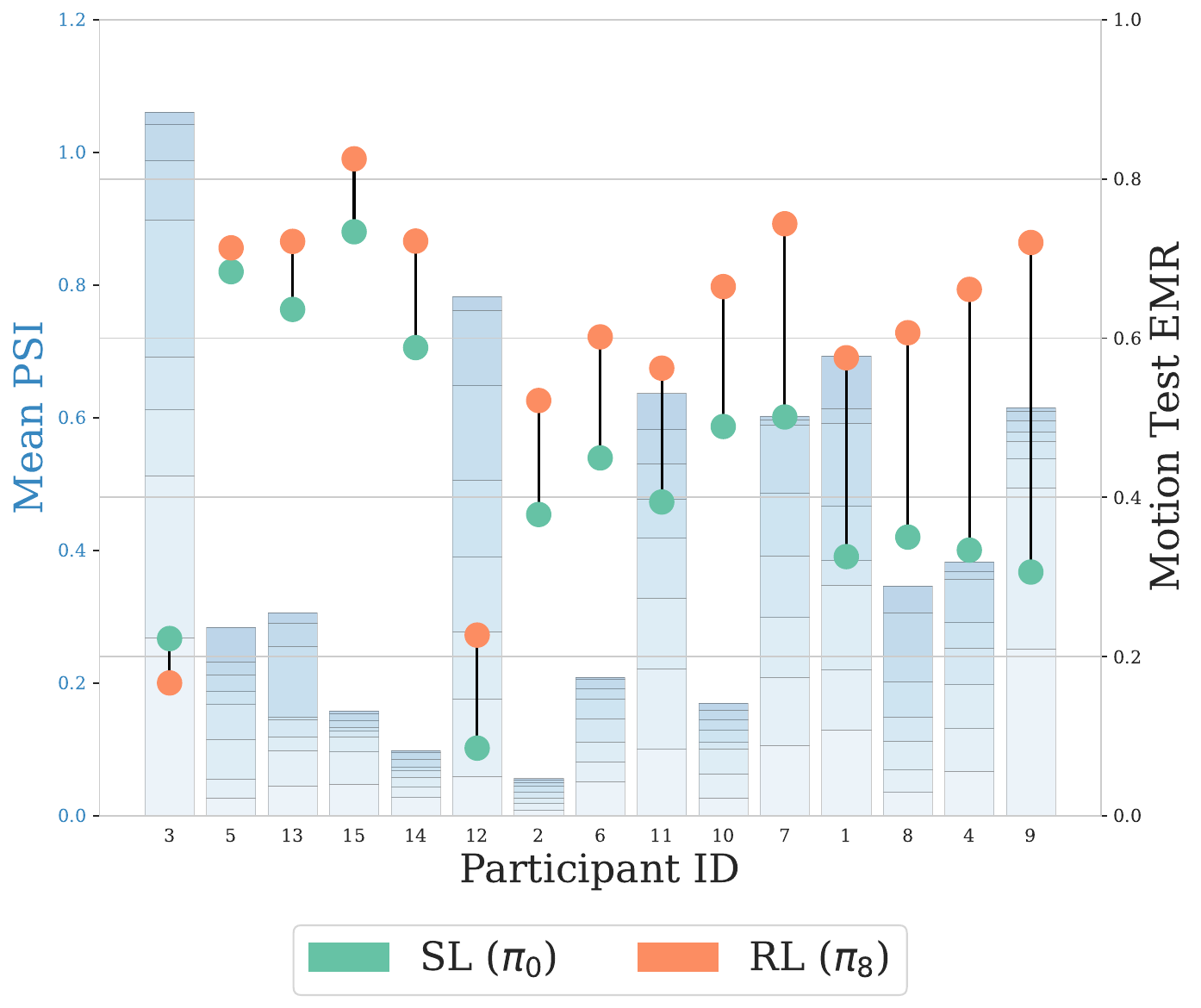}}
\caption{Mean PSI per participant between features of final episode and all previous ones. Participants are sorted by the difference in \textit{Motion Test} EMR between SL ($\pi_0$) and RL ($\pi_8$). The PSI is divided by repetitions, to display their contribution to the overall score. The repetitions are ordered from bottom to top. Note that the bar size corresponds to how different a distribution is compared to $\mathcal{S}_8)$.} 
\label{fig:entropies}
\end{center}
\vspace{-5mm}
\end{figure}

Assuming that participants become more consistent in playing the game as they get more practice (also indicated by the increase in return between repetition 0 and 9), we compute $\text{PSI}(\mathcal{S}_i, \mathcal{S}_8)$ for $ i=0,\ldots,7$. We then compute a final score as the mean of all inter-episode comparisons, leading to the PSI scores shown in Fig. \ref{fig:entropies}. The score of participant 9, which obtained the highest improvement from our method, shows the expected trend most clearly, as the PSI gradually changes less and less for each episode. This is illustrated by the decreasing size of the stacked bars with PSI scores for each repetition, starting with 0 at the bottom until 7 at the top. 

This leads us to question why our method was not able to learn for participant 3, while being successful for participant 12. We are interested in whether the feature distribution of participant 3 changed more than for others during gameplay. Results from Fig. \ref{fig:entropies} show that participant 3 had the highest mean PSI, which might have been substantial enough to prevent learning. However, participant 12 had the second highest PSI while still being able to improve, which means we cannot find a conclusive explanation based on this measure.

\subsection{Contraction force over experiment}\label{subsec:contraction}

\begin{figure*}[h]
\begin{center}
\centerline{
\includegraphics[width=\textwidth]{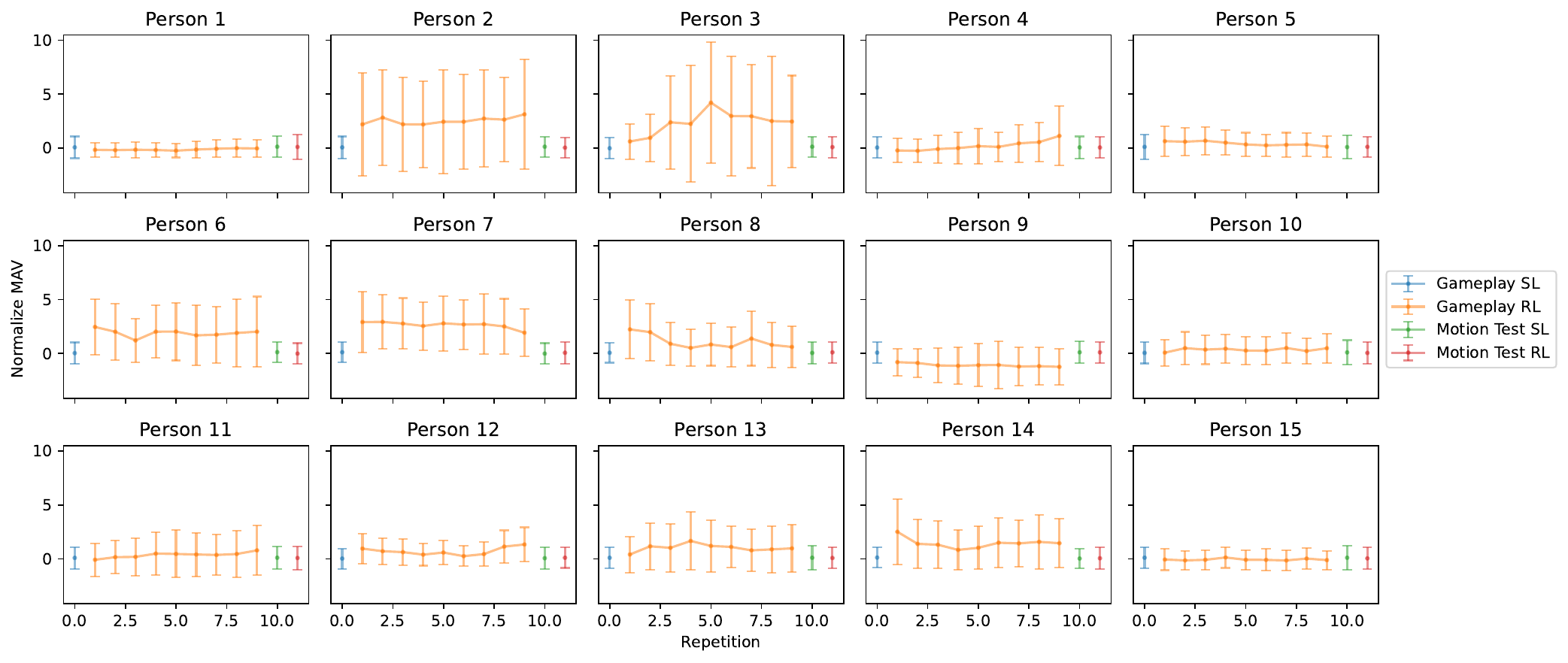}}
\caption{Mean MAV values over all channels and experiments. Pretraining data is shown in blue, RL gameplay in yellow, SL and RL \textit{Motion Tests} are shown in green and red respectively. The MAV values are normalized after collecting the pretraining data to zero mean and unit variance. Thus this figure shows the change in force in respect to the initial recordings. } 
\label{fig:mean_mavs}
\end{center}
\vspace{-5mm}
\end{figure*}

While looking for potential sources of variation in the muscle activations of the participants, we wanted to investigate how the contraction force changed over the experiments. Fig. \ref{fig:mean_mavs} shows the mean values over all channels of the MAV feature, which can be associated with the force used. The values are normalized to 0 mean and standard deviation 1 after collecting the pretraining data, in order to perform a qualitative analysis about the force change over the experiments. As can be seen, most participants contract with a relatively constant force throughout the experiments. Furthermore, it becomes clear that the used force in the \textit{Motion Tests} is much closer to the one used during pretraining, which underlines the effectiveness of our method as RL gameplay clearly improves \textit{Motion Test} performance. Notably, some participants do vary widely in their contraction level, namely participants 2 and 3. While this might give an indication of why training was unsuccessful for participant 3, it does not form a general pattern as the scores for participant 2 clearly increased despite this variation. Therefore, we argue that force is not a key factor to determine the success of our method.

\end{document}